\documentclass[useAMS,usenatbib]{mn2e}
\usepackage{graphicx}

   \title[New photometric investigation of the double ringed galaxy ESO474-G26.]{New photometric investigation of the double ringed galaxy ESO474-G26. Unveiling the formation scenario}
  \author[M. Spavone et al.]{M. Spavone$^{1,3}$\thanks{E-mail: spavone@na.astro.it (MS)}, E. Iodice$^{3}$, D. Bettoni$^{2}$, G. Galletta$^{1}$, P. Mazzei$^{2}$ and V. Reshetnikov$^{4,5}$\\
$^{1}$Dipartimento di Fisica e Astronomia, Universit\'a di Padova, Vicolo dell'Osservatorio 2, I-35122 Padova, Italy\\    
$^{2}$INAF-Astronomical Observatory of Padova, Vicolo dell'Osservatorio 5, I-35122 Padova, Italy\\
$^{3}$INAF-Astronomical Observatory of Naples, via Moiariello 16,
   I-80131 Napoli, Italy\\
          $^{4}$St.Petersburg State University, Universitetskii pr. 28, Petrodvoretz, 198504 Russia\\
   $^{5}$Isaac Newton Institute of Chile, St Petersburg Branch}
\begin{document}
\bibliographystyle{mn2e}
   \date{Accepted 2012 July 27. Received 2012 July 23; in original form 2012 June 12}

\pagerange{\pageref{firstpage}--\pageref{lastpage}} \pubyear{2012}

\maketitle

\label{firstpage}

\begin{abstract} 
We present a detailed photometric study of the peculiar double ringed
galaxy ESO474-G26. Near-Infrared (NIR) and optical data have been
used, with the main goal to constrain the formation history of
ESO474-G26. NIR photometry is fundamental in this kind of study,
because gives better constraints on the Spectral Energy Distribution
(SED) and well traces the older stellar population of the galaxy. This
galaxy presents a very complex structure, with two almost orthogonal
rings, one in the equatorial and another in the polar plane, around an
elliptical-like object. Due to the peculiar morphology of ESO474-G26,
we used both NIR images (J and K bands) to derive accurate analysis of
the stellar light distribution, and optical images (in the B, V and R
bands) to derive color profiles and color maps to study the structure
of the rings. The observational characteristic of ESO474-G26 are
typical of galaxies which have experienced some kind of interactions
during their evolution. We investigated two alternatives: a merging
process and an accretion event.
\end{abstract}

%________________________________________________________________

\section{Introduction} \label{intro}

The main aim of the extragalactic astrophysics is to understand how
galaxies formed and evolved: the advent of the new all-sky surveys,
covering a wide wavelength range, and the high resolution data from
the large ground-based and space telescopes have strongly confirmed
that gravitational interactions and mergers affect the morphology and
dynamics of galaxies from the Local Group to high-redshift universe
(\citealt{Cons03}; HDF and HUDF; SDSS). From this kind of studies,
there is a growing evidence that mergers play a major role in the
formation of early-type galaxies (Ellipticals and S0s), both in the
field and in clusters. The traditional debate about the formation of spheroids (Es)
and disk (S0s) galaxies, is today re-addressed to understand the
origin of slow and fast rotators (see \citealt{Ems11} and
\citealt{Kho11}) in which ETGs are today divided. \citet{Ems11} show
that about 66\% of elliptical galaxies are fast rotators, i.e. they
might have a disk component. \citet{Tal09} found that 73\% of nearby
ellipticals show morphological feature of interactions.

These observational results support the Cold Dark Matter scenario for
galaxy formation \citep{Cole00}: it is based on the hierarchical mass
assembly, where the observed galaxies and their dark halo were formed
through repeated mergings of small systems. In this framework, the
study of peculiar and interacting galaxies, both at low and at high
redshift, has a special role to investigate on the main processes at
work during gravitational interactions between galaxies and between
galaxies and their environment. In particular, in latest ten years, a
big effort has given to study the morphology and kinematics of Polar
Ring Galaxies (PRGs) and related objects: in these systems, the
existence of two orthogonal components of the angular momentum is a
consequence of a ``second event'' happened in their formation history,
thus, PRGs can be considered as the ideal laboratory to study both the
gravitational interactions among galaxies and the dark halo shape.

In the PRG catalogue made by \citet{Whi90}, the included objects are
all classified as polar ring galaxies, where the morphology of central
host resemble that of an early-type galaxy (Elliptical or S0) and the
polar structure is a ring made up of gas, stars and dust that orbits
in a nearly perpendicular plane with respect to central component
(\citealt{Sch83}; \citealt{Ber85}). By taking advantage of high
resolution spectroscopy and photometry, only subsequent studies on the
prototype of PRGs, NGC4650A, have revealed for the first time that the
polar structure in this object has the morphology and kinematics of a
disk, rather than a ring (see \citealt{Arn97}; \citealt{Iod02};
\citealt{Gal02}; \citealt{Swa03}). From that moment on, by comparing
observations and theoretical predictions, studies on polar ring/disk
galaxies have tried to address how different kind of interactions
(i.e. between galaxies and with environment) let to different galaxy
morphologies and kinematics. Currently, in order to account both for
the featureless morphology of the central spheroidal galaxy and for
the more complex structure of the polar ring/disk, the main formation
processes proposed are: {\it i)} a major dissipative merger; {\it ii)}
tidal accretion of material (gas and/or stars) by outside; {\it iii)}
cold accretion of pristine gas along a filament. In the merging
scenario, the PRG results from a ``polar'' merger of two disk galaxies
with unequal mass: the morphology and kinematics of the merger
remnants depends on the merging initial orbital parameters and the
initial mass ratio of the two galaxies (\citealt{Bek98a};
\citealt{Bek98b}; \citealt{Bou05}). In the accretion scenario, the
polar ring/disk may form by a) the disruption of a dwarf companion
galaxy orbiting around an early- type system, or by b) the tidal
accretion of gas stripping from a disk galaxy outskirts, captured by
an early-type galaxy on a parabolic encounter (\citealt{Res97};
\citealt{Bou03}; \citealt{Han09}). The cold accretion scenario has
been proposed very recently for the formation of a wide disk-like
polar rings: a long-lived polar structure may form through cold gas
accretion along a filament, extended for $\sim 1$ Mpc, into the
virialized dark matter halo (\citealt{Mac06}; \citealt{Brook08}). In
this formation scenario, there is no limits to the mass of the
accreted material, thus a very massive polar disk may develop either
around a stellar disk or a spheroid. From the ``observational'' side,
as suggested by very recent studies on PRGs (\citealt{Iod06};
\citealt{Spav10, Spav11}), the critical physical parameters
that allow to discriminate among the three formation scenarios are 1)
the total baryonic mass (stars plus gas) observed in the polar
structure with respect to that in the central spheroid; 2) the
kinematics along both the equatorial and meridian planes; 3) the
metallicity and SFR in the polar structure. By studying the chemical
abundances in the polar structure of three polar disk galaxies,
NGC4650A, UGC7576 and UGC9796, \citet{Spav10, Spav11} have
traced the formation history of these objects by accounting for all
the three parameters mentioned above. In particular, the cold
accretion scenario was successfully tested for the first time.

In the present paper, we address the formation history of the multiple
ring galaxy ESO474-G26, by comparing the observed structure with the
predictions from different formation scenarios. Together with the
previous one (cited above), this work is part of an ongoing research
project which aim to study the morphology, kinematics and SFR of a
statistically significant sample of polar ring/disk galaxies and
related objects, selected from both the Whitmore's PRG catalogue and
from the new PRG catalogue compiled by \citet{Moi11} based on
SDSS data.

\subsection{Properties of the PRG ESO474-G26}

\begin{figure*}
\centering
\includegraphics[width=12cm]{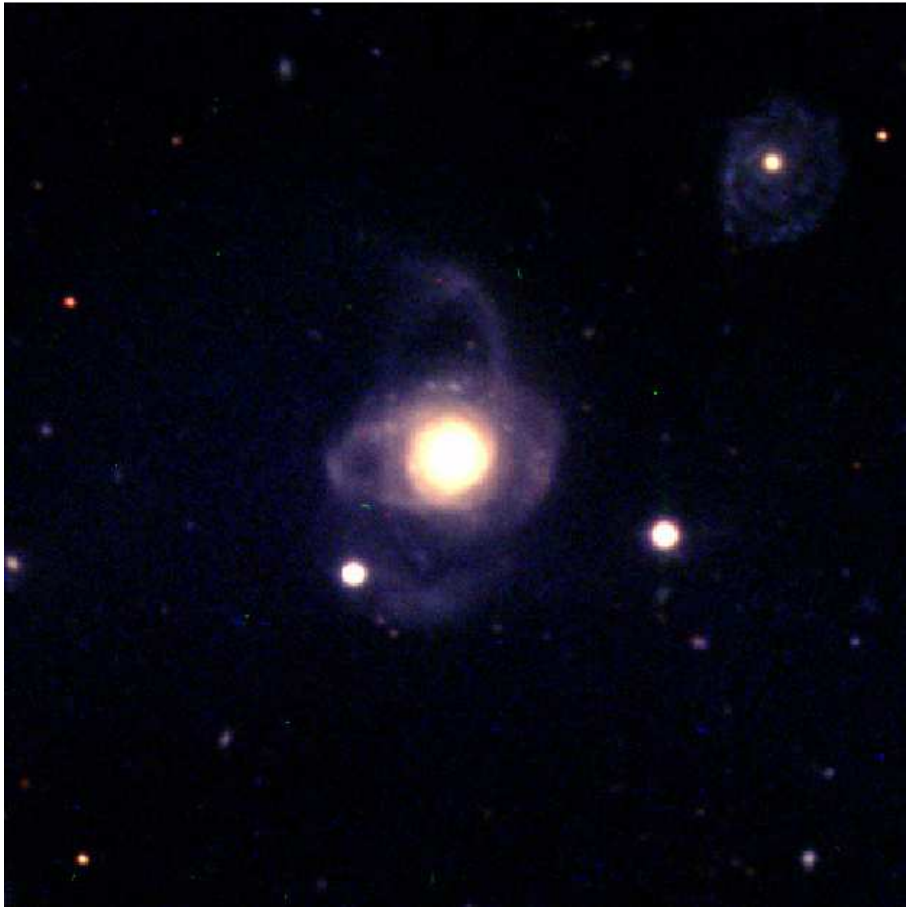}
\caption{Color composite image of ESO474-G26 assembled from images in
  the B (blue channel), V (green channel) and R (red channel)
  bands. The north is up, while the east is on the left of the
  image.} \label{ESO474}
\end{figure*}

ESO474-G26 (Figure \ref{ESO474}) has been classified as a \emph{possible
  candidate for polar ring galaxy} (PRC C-3) by \citet{Whi90}. This
object has a heliocentric radial velocity of $V = 15802\ km \ s^{-1}$,
which implies a distance of about 211 Mpc, based on $H_{0} = 75 \ km
\ s^{-1} \ Mpc^{-1}$, and with this distance 1 arcsec $\simeq$ 1 kpc.

This galaxy has two perpendicular and almost irregular rings, one in
the equatorial and one in the polar plane, surrounding a central
nearly spherical galaxy. \citet{Res05} found very blue optical colors
for the rings, typical of late type spirals. Moreover, since both
rings rotate around the central galaxy (\citealt{Whi90};
\citealt{Res05}), they conclude that ESO474-G26 can be classified as a
kinematically confirmed PRG.

Optical spectra also show that the ionized gas rotates with the north
and west side receding. \citet{Gal97} by analyzing the relatively
strong CO signal in ESO474-G26 found that the outer ring (north-south)
has the northern side receding, indicating that the molecular gas rotates
in the same direction as the ionized one.

The field around ESO474-G26 does not show any object within 10
arcmin or 600 kpc, since the galaxy visible on the NW side is a
background object \citep{Gal97}.

\citet{Res05} analyzing the luminosity profiles of ESO474-G26 in the
optical bands were able to distinguish three components: \emph{i}) the
main central body with almost round isophotes ($b/a \sim 0.94$),
\emph{ii}) a narrow ring, with a diameter of $\simeq 37''=37$
kpc, in the equatorial plane, and \emph{iii}) a second (larger) polar
ring, with a diameter of $\simeq 58''=58$ kpc, in the polar
plane. Both rings are very irregular.

They also found that the galaxy colors, the ratio of the $H_{2}$ mass
to blue luminosity and the HI content, correspond to those of an
Sb-Sbc spiral, even if the Spectral Energy Distribution (SED) for a
prototype advanced merger remnant well fit the observed SED for
ESO474-G26. The central galaxy appears to be redder than the galaxy as
a whole, but bluer than typical ellipticals, while both rings are
bluer than the central body and have colors typical of PRG rings
(\citealt{Res94p, Res95}).

The main properties of ESO474-G26 are listed in Table
\ref{global}.

\begin{table*}
\begin{minipage}[t]{\columnwidth}
\caption{Global properties of ESO474-G26.}
\label{global}
 \centering
\renewcommand{\footnoterule}{}     
\begin{tabular}{lccc}
\hline\hline
Parameter&Value&Ref.\\
\hline
Morphological type&Sc peculiar&NED\footnote{NASA/IPAC Extragalactic Database}\\
R.A. (J2000)           &00h47m07.5s&NED \\
Decl. (J2000)           &-24d22m14s&NED \\
Helio. radial velocity &15802 km/s&NED\\
Redshift         &0.052710 &NED\\
Distance      &211 Mpc     & \\
Axial ratio& 0.65& Reshetnikov et al. (2005)\\
Absolute magnitude $M_{B}$& -22.04&Reshetnikov et al. (2005)\\
Absolute magnitude $M_{J}$& -24.67&This paper\\
Absolute magnitude $M_{K}$& -25.52&This paper\\
$M(HI)(M_{\odot})$  & $1.7 \times 10^{10}$&Reshetnikov et al. (2005)\\
$M(H_{2})(M_{\odot})$  & $2.3 \times 10^{10}$&Galletta et al. (1997)\\
$M(HI)/L_{B}(M_{\odot}/L_{\odot})$ &0.15& Reshetnikov et al. (2005)\\
$M(H_{2})/L_{B}(M_{\odot}/L_{\odot})$ &0.20& Reshetnikov et al. (2005)\\
\hline
\end{tabular}
\end{minipage}
\end{table*}

\subsection{Radio emission and star formation rate}\label{radio}

\begin{figure*}
\centering
\includegraphics[width=15cm]{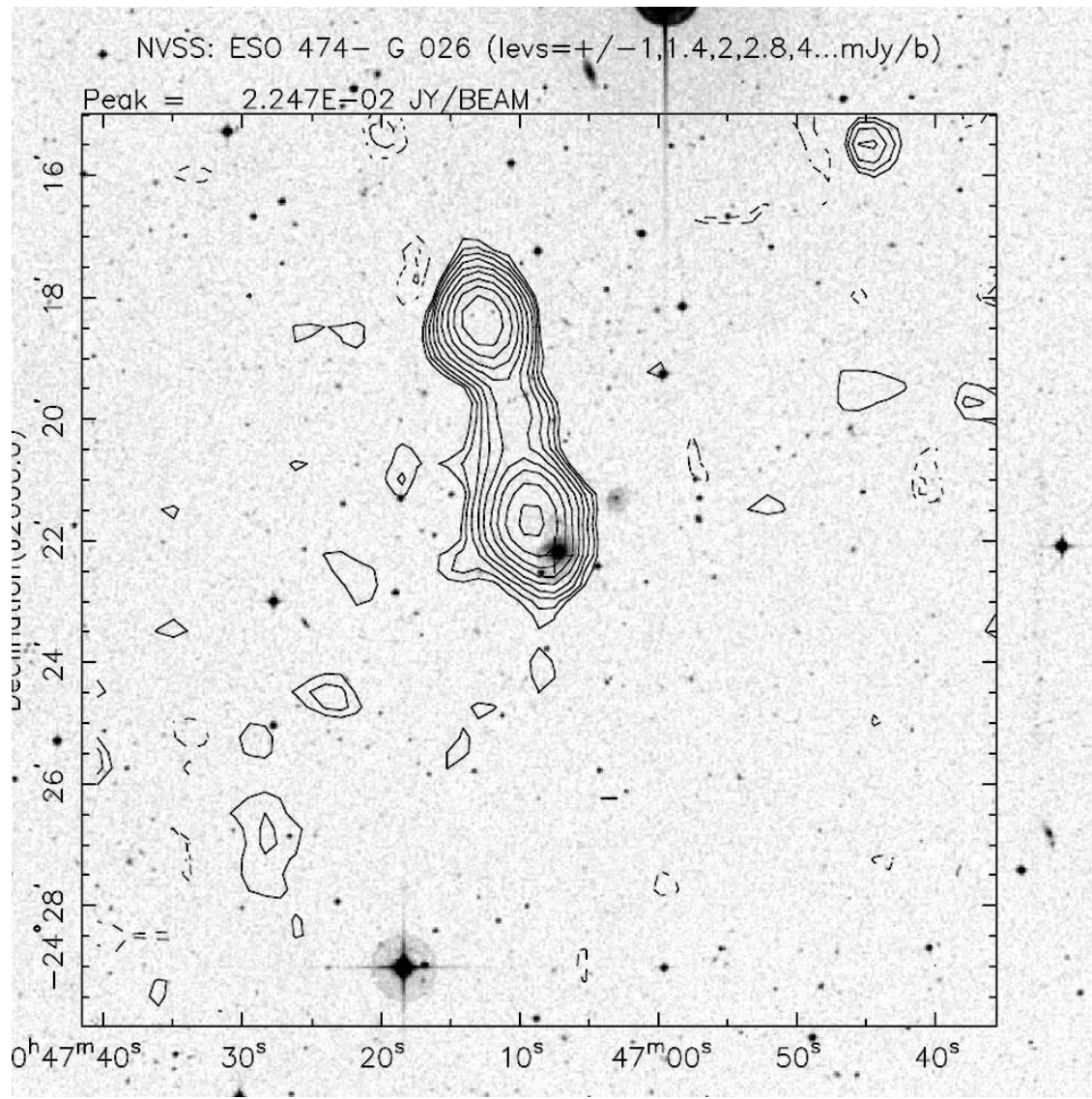}
\caption{21 cm contours superimposed to the B band image of
  ESO474-G26. The north is up, while the east is on the left of the
  image.} \label{21cm}
\end{figure*}

In Figure \ref{21cm} is shown the POSS image of the galaxy with
the isocontours of the 1.4 GHz continuum emission from the NRAO VLA
Sky Survey (NVSS, \citealt{Con98}) superimposed. The radio emission is
elongated in the Northern direction and \citet{Con98} reported
three sources (the two peaks and the flux in between). If we consider
only the emission closest to ESO474-G26 the flux is F$_{1.4}$ =
4.65$\times10^{-28}$ W $m^{-2}Hz^{-1}$ corresponding, with our adopted
distance (see Table \ref{global}), to a luminosity L$_{1.4}$=
2.43$\times10^{23}$W$Hz^{-1}$. The 1.4 GHz luminosity is insensitive
to dust obscuration and for this reason is a good tracer of the star
formation rate (SFR1.4). We adopt the calibration of \citet{Hop03} and
we found a SFR=130$M_{\odot} y^{-1}$ (see Table \ref{nvss}).

\citet{Res05}, converting the far-infrared luminosity to a star
formation rate, also found a high rate of star formation
($SFR_{FIR} = 43 M_{\odot}/yr$).

\begin{table*}
\caption{\label{nvss}Radio fluxes, luminosities and SFRs of the three sources detected by NVSS.} \centering
\begin{tabular}{lccccccc}
\hline\hline
NVSS & RA (J2000)& Dec (J2000)& S1.4 (mJy)& err(mJy)&$L_{1.4}$ (W/Hz) & SFR ($M_{\odot}$/yr)\\
\hline
004709-242140 & 00 47 09.06 & -24 21 40.4 & 46.5 & 2.4& 2.4291e+23 &   1.3e+2\\
004710-241949 & 00 47 10.51 & -24 19 49.3 & 9.1 & 1.3& 4.8475e+22  &  26.8\\
004712-241823 & 00 47 12.67 & -24 18 23.5 & 40.9 & 2.0& 2.1787e+23 &   1.2e+2\\
\hline
\end{tabular}
\end{table*}
%__________________________________________________________________

\section{Observation and data reduction} \label{data}

{\it Near-Infrared data} - ESO474-G26 belongs to a selected sample of
peculiar galaxies observed in the Near-Infrared (NIR) J and K bands in
December 2002, with the SofI infrared camera at the ESO-NTT
telescope. The field of view was $4.92 \times\ 4.92\ arcmin^{2}$ with a
pixel scale of 0.292 arcsec/pixel. Images were acquired in the
offsetting mode: a cycle was defined by several images on the target,
interspersed with sky frames and with an integration time of 60
seconds; each object frame was taken with a small offset from the
galaxy center and the sky frames were taken before and after each
galaxy frame. More cycles were obtained in the K band than in the J
band, in order to have a better estimate of the background level. A
total exposure time of 360 sec was obtained on the target in the J
band and of 1080 sec in the K band. The average seeing during the
observing time is about FWHM $\simeq$ 1.1 arcsec.

The data reduction was carried out using the {\small CCDRED} package
in the IRAF\footnote{IRAF is distributed by the National Optical
  Astronomy Observatories, which is operated by the Associated
  Universities for Research in Astronomy, Inc. under cooperative
  agreement with the National Science Foundation.}({\it Image
  Reduction and Analysis Facility}) environment. The main strategy
adopted for each data-set included dark subtraction\footnote{Bias
  frame is included in the Dark frame.}, flatfielding correction, sky
subtraction and rejection of bad pixels. Finally, all frames were
registered and co-added to form the final science frames.

Several standard stars, from \citet{Per98}, observed at the beginning,
middle and end of each observing night, were used to transform
instrumental magnitudes into the standard J and K band systems. The
obtained photometric zero points are $Z_{P}(J) = 23.04 \pm 0.02\
mag/arcsec^{2}$ for the J band and $Z_{P}(K) = 22.35 \pm 0.02\
mag/arcsec^{2}$ for the K band.

\begin{figure*}
%\centering
\includegraphics[width=8cm]{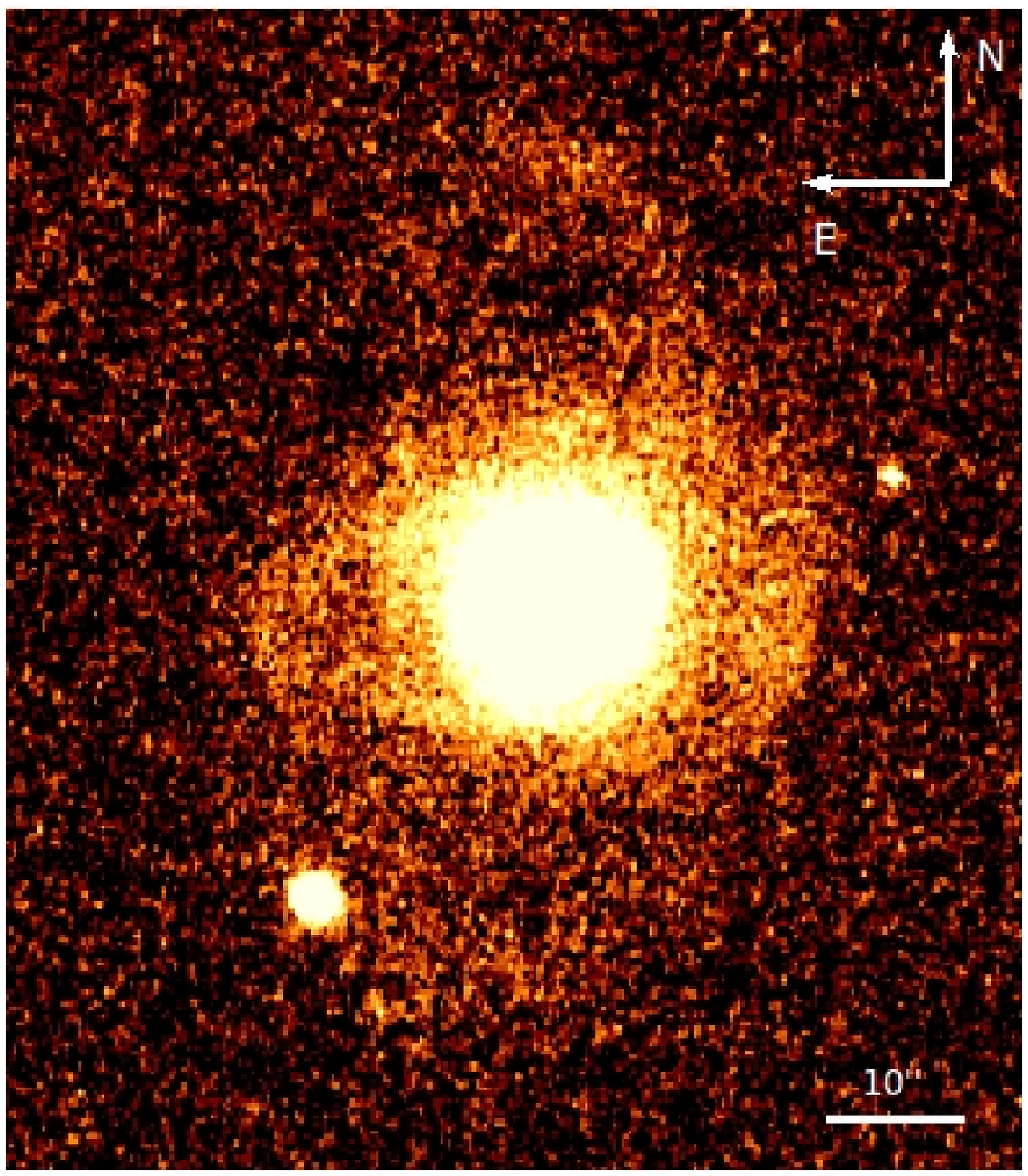}
\includegraphics[width=8cm]{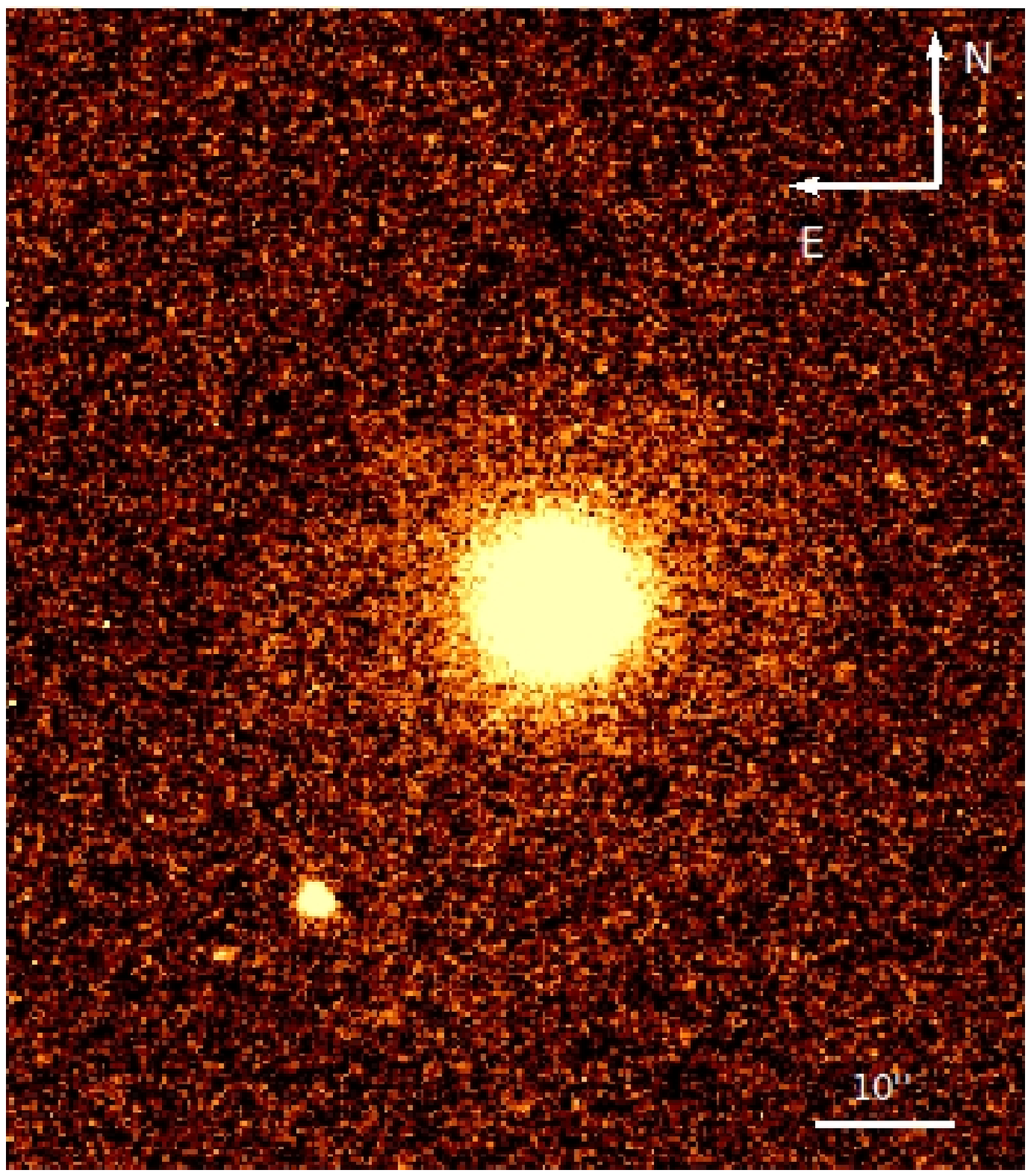}
\caption{{\it Left
    panel}-J-band image of ESO474-G26. {\it Right panel}-K-band image of ESO474-G26.} \label{ESO474JK}
\end{figure*}

The calibrated J and K band images of ESO474-G26 are shown in
Figure \ref{ESO474JK}. The ring-like structure of the galaxy is still
visible in the J band image, while it disappears in the K band one.

{\it Optical data} - Photometric observations in the Johnson B, V and
Cousins R bands were obtained in August 2002 on the 1.6 m telescope of
the Observatorio do Pico dos Dias (operated by the MCT/Laboratorio
Nacional de Astrofisica, Brazil), equipped with direct imaging camera
and a CCD detector with a pixel scale of 0.18 arcsec/pixel. The
average seeing during the observing time is about FWHM $\simeq$ 1.3
arcsec. Reduction of the CCD frames was performed as described in
  \citet{Res05}. The photometric calibration was made by using
  standard stars from the \citet{Lan83} and \citet{Gra82} lists,
  obtaining the following photometric zero points: $Z_{P}(B) = 23.29
  \pm 0.06\ mag/arcsec^{2}$, $Z_{P}(V) = 23.23 \pm 0.06\ mag/arcsec^{2}$
  and $Z_{P}(R) = 23.26 \pm 0.06\ mag/arcsec^{2}$, for the B, V and R
  bands respectively.

\section{Host galaxy and rings morphology}\label{morph}

The NIR images of ESO474-G26 (Figure \ref{ESO474JK}) show that most of
the NIR light comes from the host galaxy and its morphology
resembles that of an almost round elliptical object. The equatorial
ring, with a diameter of about 40 kpc, is within the optical radius of
the central galaxy ($\sim$ 40 kpc), while the polar one has a diameter
of about 60 kpc and so it is more extended in radius than the host
galaxy.  Both rings are more clearly visible in the optical B band
image, while they gradually disappear in the J and K bands. However,
both NIR and optical images show that the host galaxy is the dominant
luminous component, while the rings appear knotty and dusty.

\begin{figure*}
\centering
\includegraphics[width=10cm]{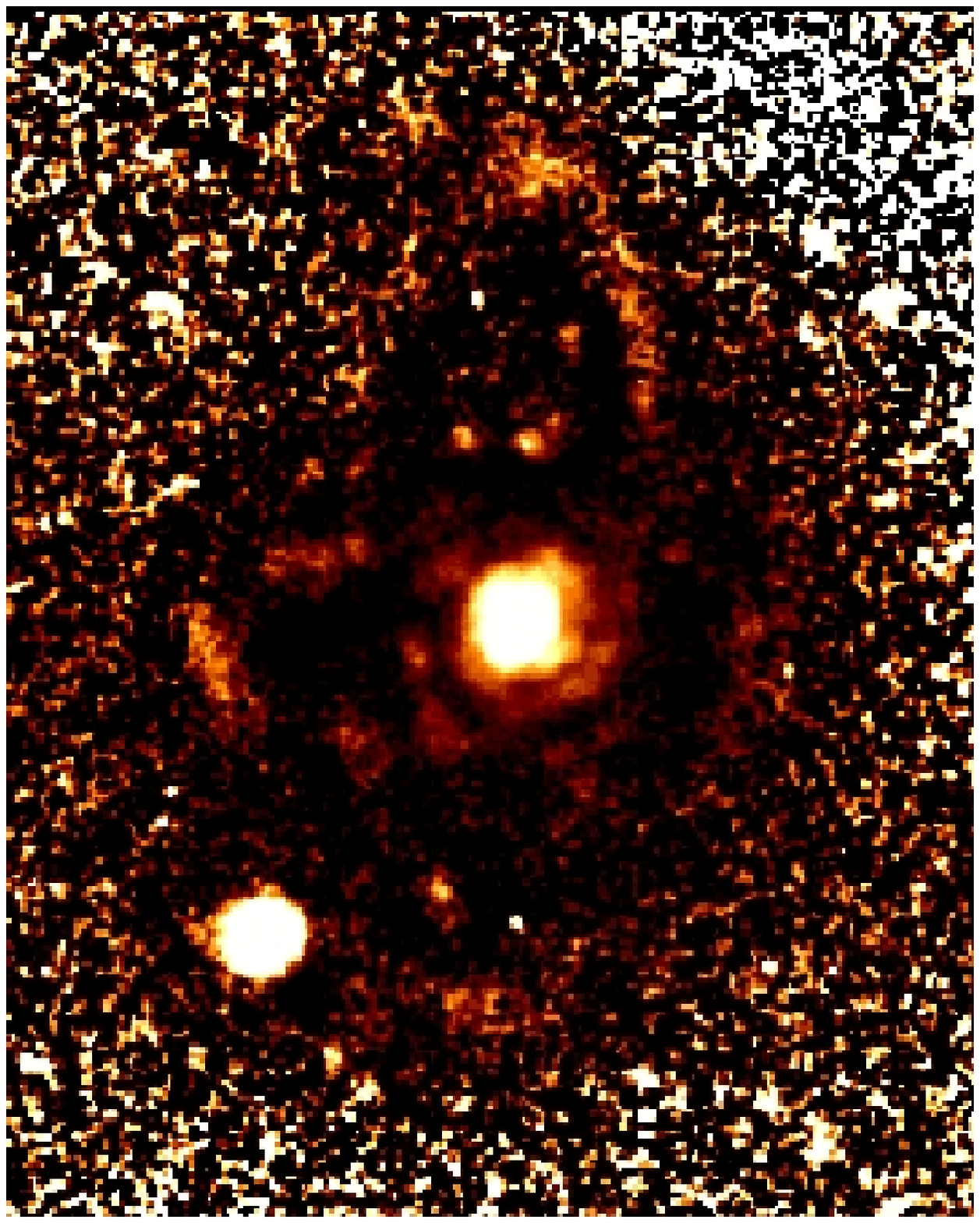}
\caption{High frequency residual B band image for ESO474-G26. Lighter
  colors correspond to brighter features. The image size is $90''
  \times 115''$, and the north is up while the east is on the
  left.} \label{ratio}
\end{figure*}

To examine the inner structure of the central host galaxy, and to
identify the high frequency residuals with respect to the homogeneous
light distribution, we create a residual image produced by taking the
ratio of the original reduced image with a smoothed one, where each
original pixel value is replaced with the median value in a
rectangular window. This has the effect of remove the large-scale
structure in the image and emphasize the galaxy substructure. We use
the IRAF task {\small FMEDIAN} to smooth the original reduced image,
by using a two-dimensional window. The window size ($7\times 7$) is
chosen to best emphasize the inner structure of the central host. The
final un-sharp masked image is shown in Figure \ref{ratio} and it
represents the {\it high frequency residual image} of ESO474-G26.

The most important result obtained by this analysis is the absence of
any disk-like structures associated with the host galaxy major
axis. The absence of a disk in the host galaxy suggested the use of a
Sersic law for the 2D fit of the light distribution in this component
(see Sec. \ref{2D}).

\section{Photometry: Light and color distribution}\label{phot}

The overall morphology of ESO474-G26 is very tricky, due to the
presence of two, almost perpendicular, ring-like structures. NIR
photometry is necessary to reduce as much as possible the dust
absorption that affect the starlight distribution and to accurately
analyse it as well as to easily identify the inner structure of
ESO474-G26. In addition, optical images are used to derive optical
versus NIR color profiles and color maps to study the peculiar
structure of this galaxy.

\subsection{Isophotal analysis} 
\begin{figure*}
\centering
\includegraphics[width=12cm]{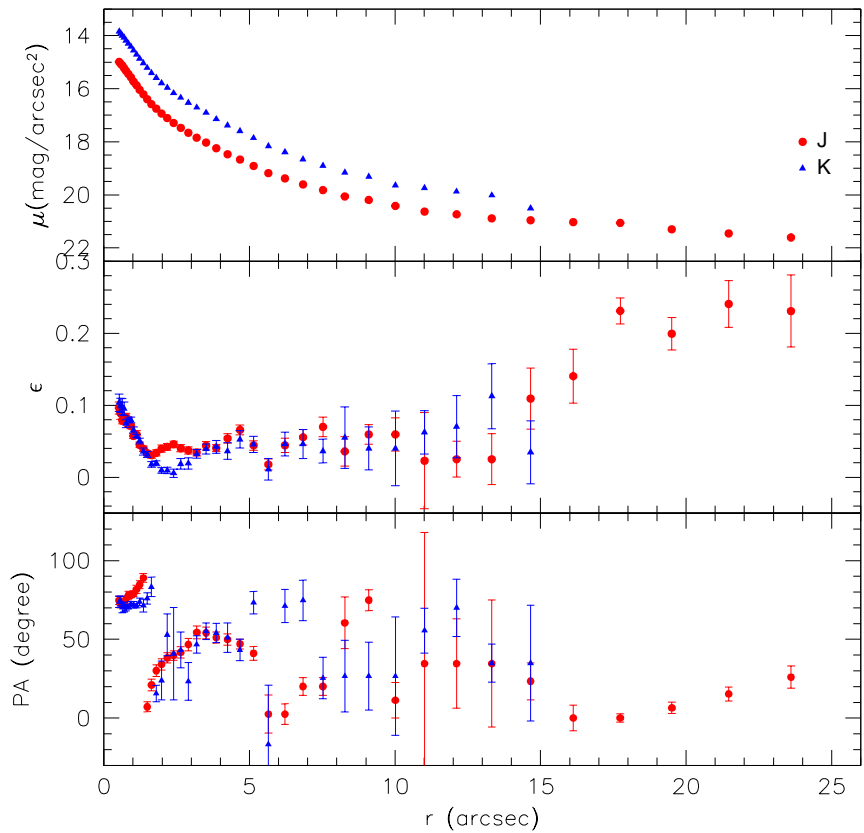}
\caption{Position Angle (P.A.), Ellipticity ($\epsilon$) and mean
  surface brightness profile in the J and K bands. The error bar for
  the surface brightness profile ($\pm$ 0.02) is within the dimensions of data
  points.} \label{photJK}
\end{figure*}

\begin{figure*}
\centering
\includegraphics[width=15cm]{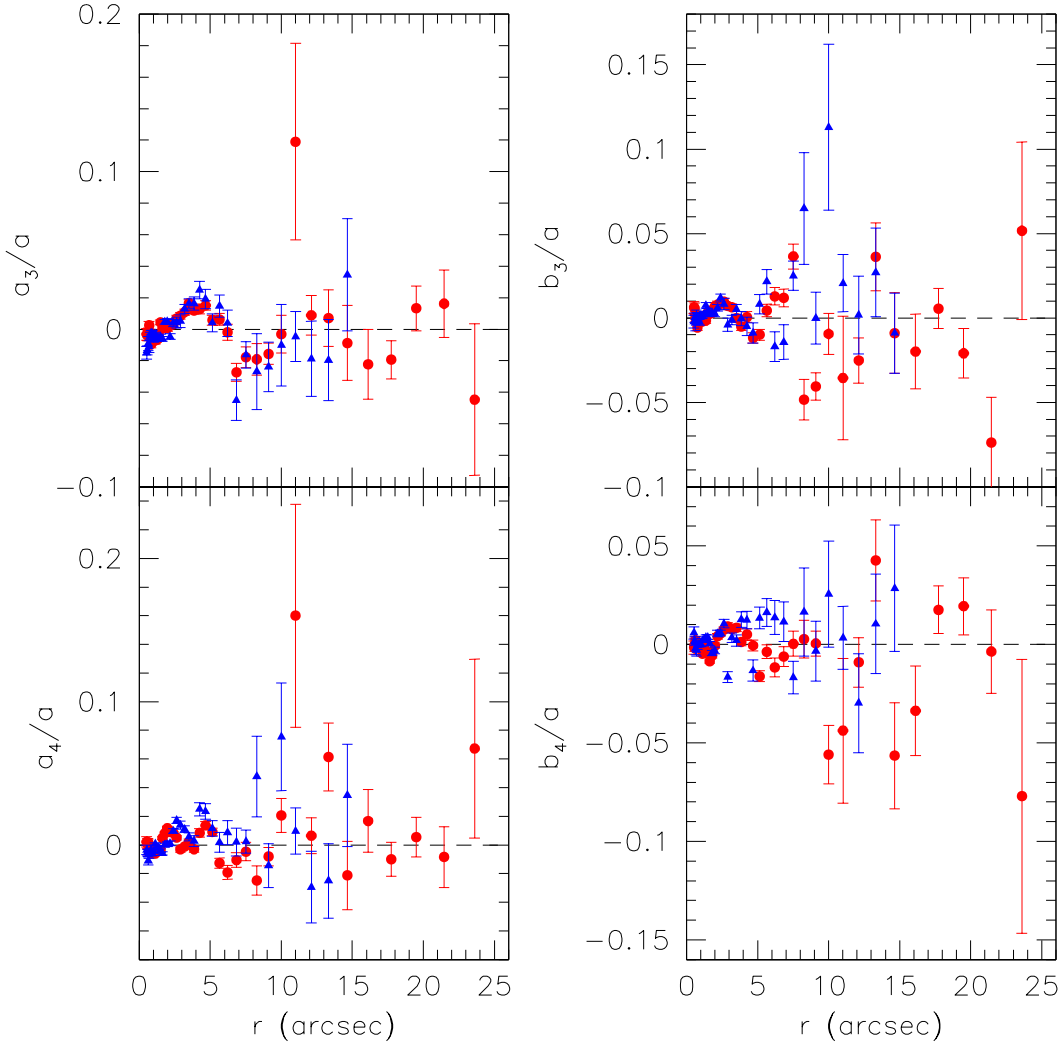}
\caption{Shape parameters in the J and K bands.} \label{a4aJK}
\end{figure*}

We used the {\small IRAF-ELLIPSE} task on the NIR images to perform
the isophotal analysis for ESO474-G26 and the results are shown in
Figure \ref{photJK}. The average surface brightness extends up to about
25 and 15 arcsec from the galaxy center for the J and K band
respectively; in the K band the half-light radius is $R_{e}=9.3$
arcsec, while in the J band is $R_{e}=10.5$. For a semi-major axis r,
in the range $2\leq r \leq 15$ arcsec, the ellipticity and the
Position Angle (P.A.) are almost constant and equal to 0.05 and $\sim
40^{\circ}$, that indicates that in this regions the isophotes are
almost round and coaxial. For $0\leq r \leq 2$ arcsec the ellipticity
shows the presence of a flatter structure in the center, with a P.A. of
$\sim 80^{\circ}$ and a twisting of the isophotes of about 50
degrees. For $r \geq 15$ the profiles for the J band result perturbed
by the presence of the rings. The shape parameters (Figure \ref{a4aJK})
are all consistent with zero, thus the isophotes do not significantly
deviate from purely elliptical shape.

\begin{figure*}
\centering
\includegraphics[width=12cm]{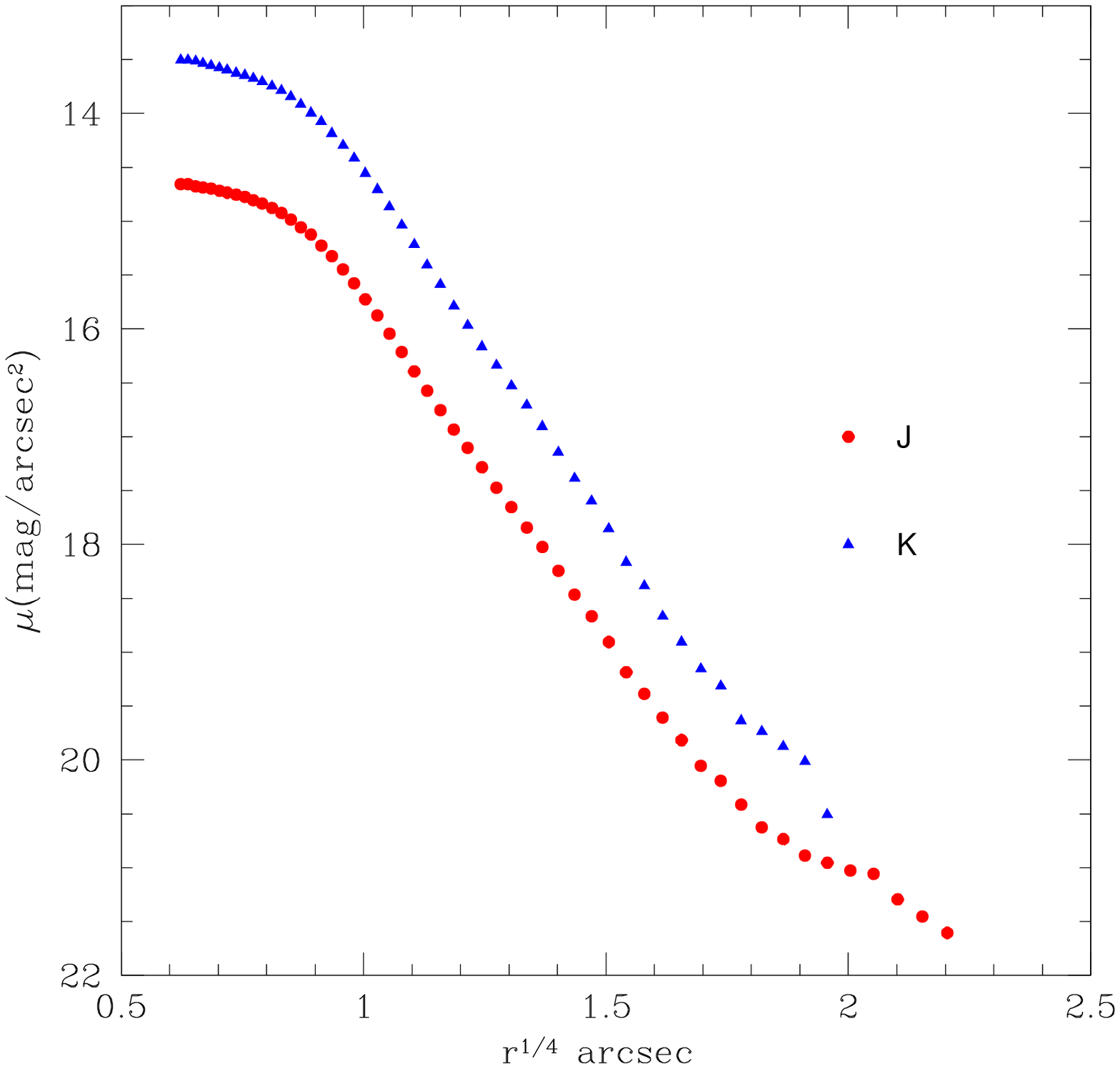}
\caption{de Vaucouleur radial surface brightness profile. The error
  bar ($\pm$ 0.02) is within the dimensions of data points.} \label{r1/4}
\end{figure*}

Figure \ref{r1/4} shows that the radial surface brightness between
  $1''$ and $10''$ is well reproduced by a de Vaucouleur profile, while in
  the outer regions we observe a bump, in both J and K profiles, which
  reflect the presence of the ring-like structures and clearly stands
  out also in the light profiles (see Figure \ref{profmodK}).

In Table \ref{mag} we give the total integrated magnitudes within two
circular apertures, derived for the NIR J and K bands. The apertures
were chosen in order to make easier the comparison with the magnitudes
of 2MASS data.

\begin{table*}
\begin{minipage}[t]{\columnwidth}
\caption{\label{mag}Magnitudes for ESO474-G26 in circular apertures.} \centering
\renewcommand{\footnoterule}{} 
\begin{tabular}{lccccc}
\hline\hline
Aperture radius& $m_{J}$ & $m_{K}$ & $m_{J} (2MASS)$ & $m_{K} (2MASS)$ \\
(arcsec)& $\pm 0.02$ & $\pm 0.02$ & -\footnote{The errors on 2MASS magnitudes are 0.01,
    for both J and K, for the 14 arcsec aperture, and 0.02 and 0.04,
    respectively for J and K, for the 24 arcsec aperture.} & - \\
\hline
14.7 & 12.16 & 11.10 & 12.40 & 11.30 \\
24.0 & 11.95 & - & 12.23 & 11.16 \\
\hline
\end{tabular}
\end{minipage}
\end{table*}

\subsection{Color distribution and integrated magnitudes}\label{colors} 
\begin{figure*}
\centering
\includegraphics[width=10cm]{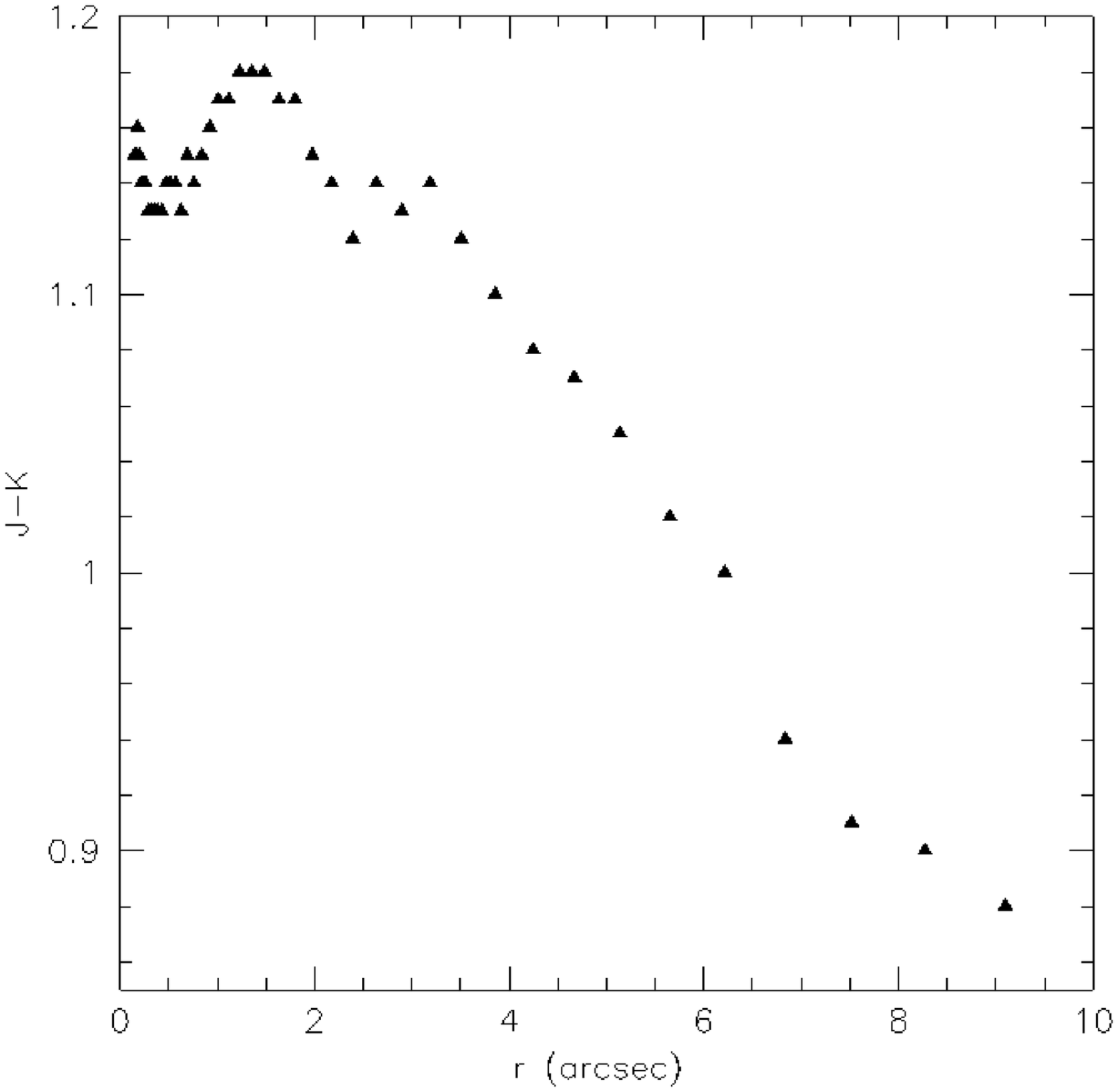}
\caption{Mean J-K color profile. The error
  bar ($\pm$ 0.04) is within the dimensions of data points.} \label{J-K}
\end{figure*}

\begin{figure*}
%\centering
\includegraphics[width=9cm]{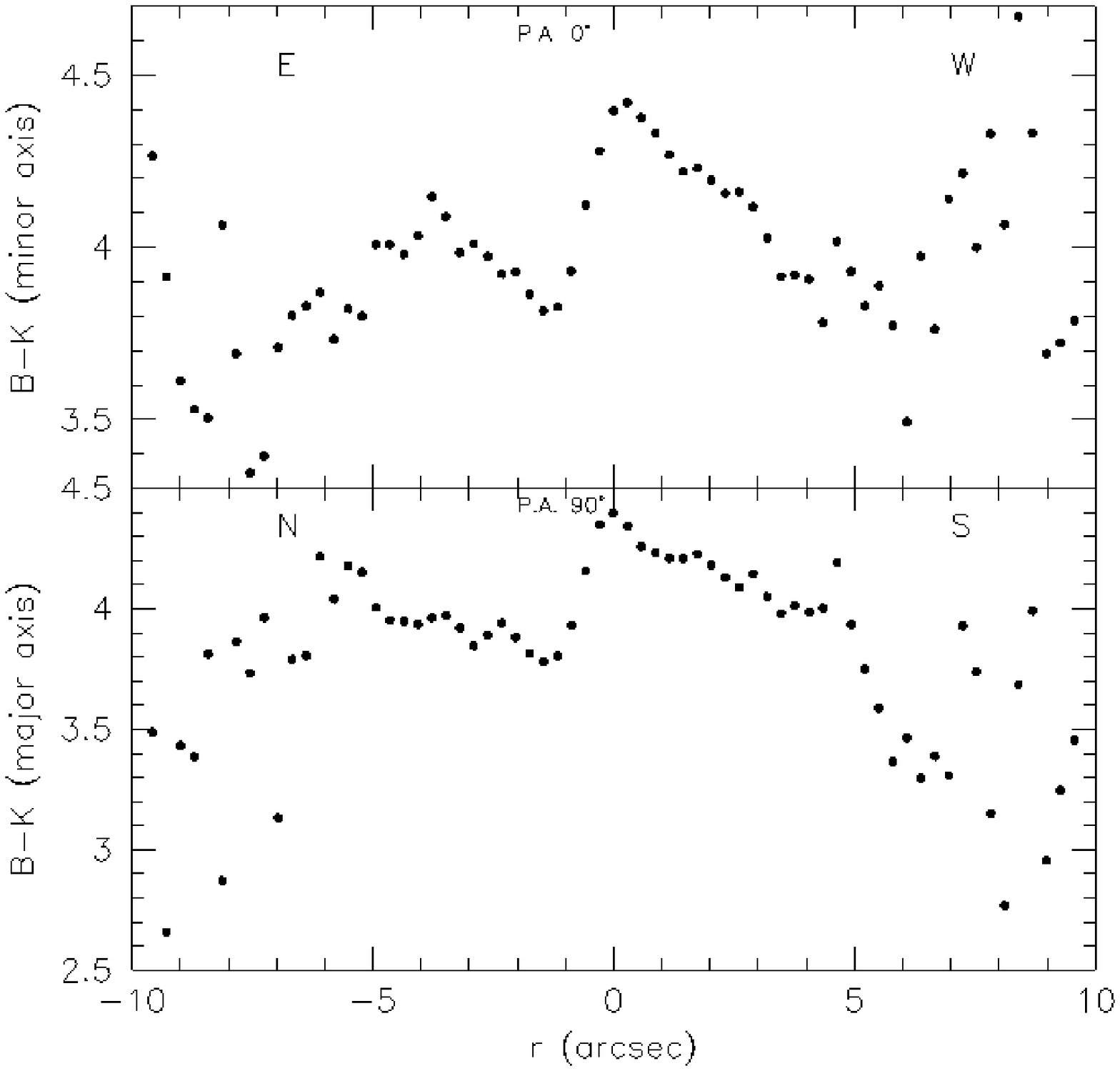}
\includegraphics[width=8cm]{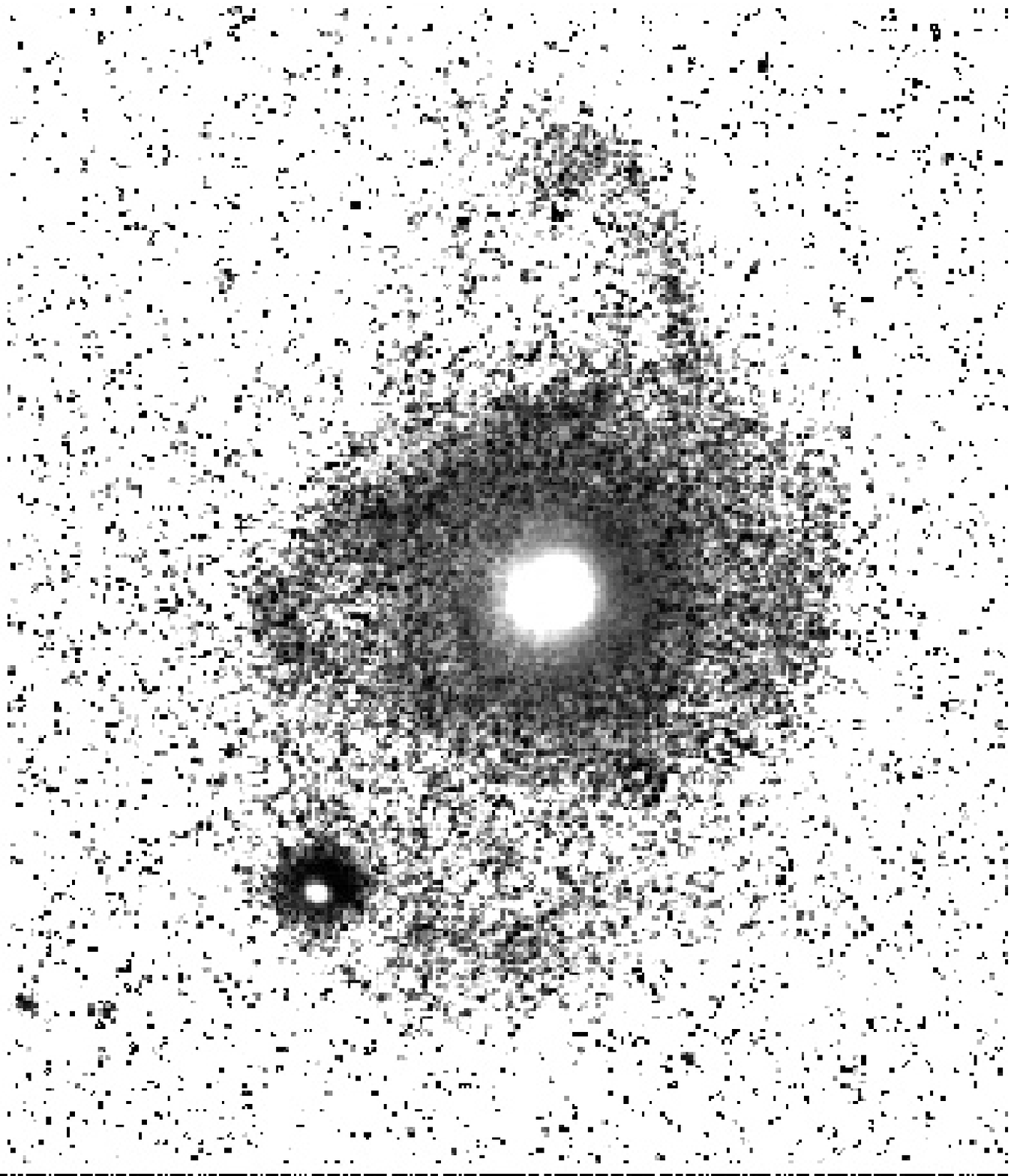}
\caption{Left panel - B-K color profiles along the minor (top panel)
  and major (bottom panel) axis. The error bar ($\pm$ 0.08) is within
  the dimensions of data points. Right panel - B-K color map. The
  North is up, while the east is on the left of the image. Lighter
  colors correspond to redder galaxy regions.} \label{B-K}
\end{figure*}

We have derived the mean J-K color profile (Figure \ref{J-K}), and B-K
(Figure \ref{B-K}, left panel) color profiles along both photometric
axes of ESO474-G26, and the 2-dimensional B-K color map
(Figure \ref{B-K} right panel). On average,
the central regions of the galaxy have redder colors, with a maximum
value of J-K $\sim 1.14 \pm 0.04$ and B-K $\sim 4.50 \pm 0.08$. As
already showed by the un-sharp masked image, also in the 2D B-K color map
we find no trace of a disk-like structure.

\begin{figure*}
\centering
\includegraphics[width=10cm]{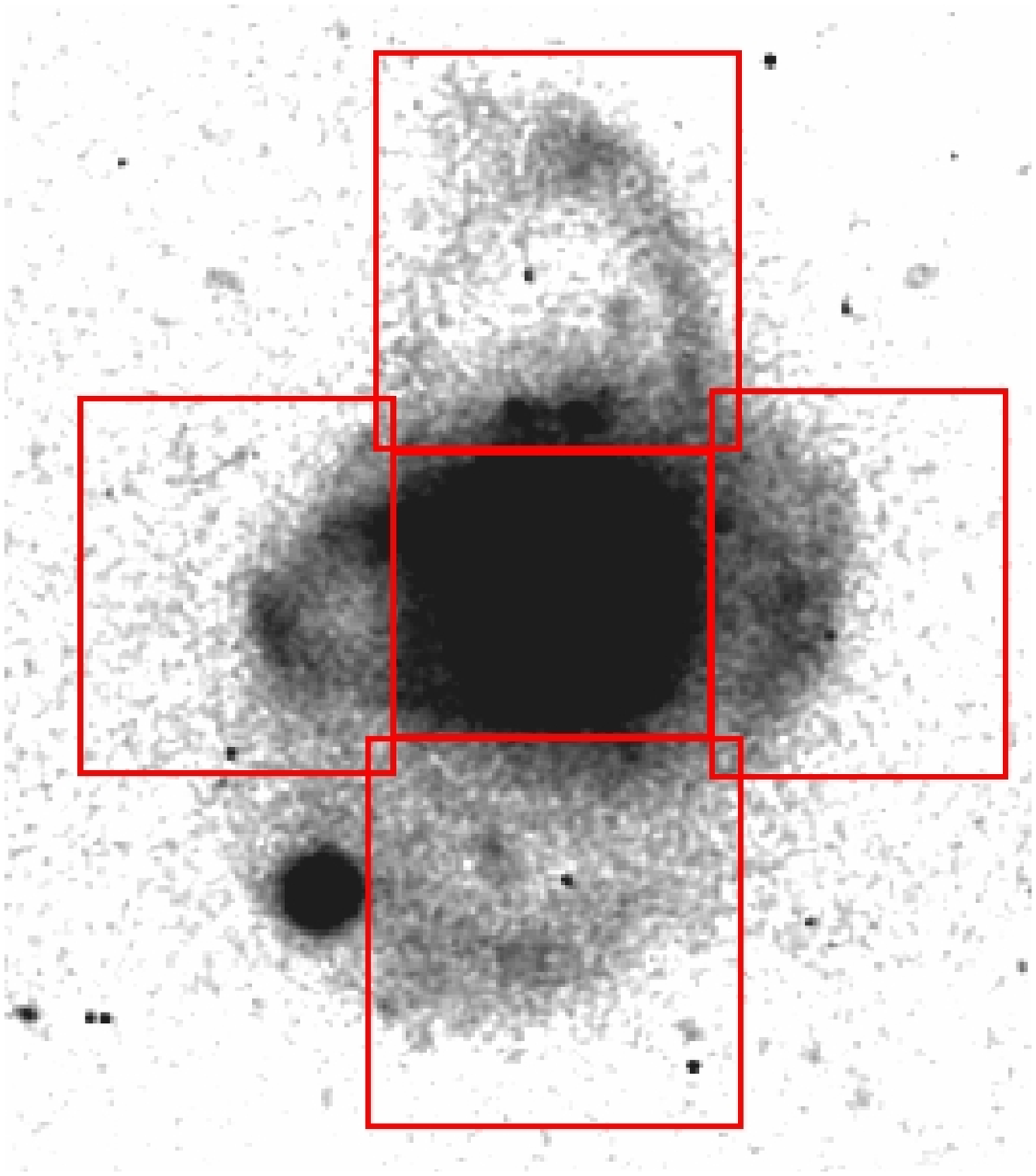}
\caption{B band image with superimposed the five rectangles limiting the
  different areas where the integrated magnitudes have been
  computed. The North is up, while the east is on the left of the
  image.} \label{poligoni}
\end{figure*}

We also derived the integrated magnitudes and J-K and B-K colors in 5
rectangles, as shown in Figure \ref{poligoni}: one including the central
region of ESO474-G26 and 4 including different regions of the
rings. The rectangles are determined from the B band image, using the
IRAF task {\small POLYMARK}, and used for all bands after the images
were registered and scaled. The integrated magnitudes inside each
rectangle are evaluated using the IRAF task {\small POLYPHOT}.

The derived magnitudes and colors are reported in
Table \ref{magnitudes}.

\begin{table*}
\caption{\label{magnitudes}Integrated and absolute magnitudes and colors of different regions of ESO474-G26.} 
\centering
\begin{tabular}{lcccccccccc}
\hline\hline
Component& Region & $m_{B} (mag)$ & $m_{J} (mag)$ & $m_{K} (mag)$ & $M_{B}$ & $M_{J}$ & $M_{K}$ & B-K & J-K\\
         &        & $\pm 0.06$   & $\pm 0.02$   & $\pm 0.02$    &        & & & $\pm 0.08$& $\pm 0.04$\\
\hline
HG & center & 13.94 & 12.51 & 11.39 &-22.68 & -24.11 & -25.23 & 2.55 & 1.12\\
PR & W & 15.30 &14.09 & 13.27 &-&-&-& 2.03 & 0.82\\
PR & E & 16.67 &15.24 & 14.78 &-&-&-& 1.89 & 0.46\\
PR & N & 15.50 & 14.16 & 13.39 &-&-&-& 2.11 & 0.77\\
PR & S & 16.97 & 15.45 & 14.99 &-&-&-& 1.98 & 0.46\\
\hline
\end{tabular}
\end{table*}

\section{Using colors to date the stellar population}\label{age}
We analyze the integrated colors (optical vs NIR) derived for the
rings and spheroid in ESO474-G26 in order to date the average
stellar populations of these main components. Taking into account
that the integrated colors are the result of both old and young
stellar populations, by studying them one can only obtain an
indication on how much one is more prominent than the other in the two
galaxy components, i.e. Host Galaxy (HG) and rings. As a consequence, the age
estimate is the average value relative to all stellar populations
present, which is strongly biased by the last burst of star
formation. In the case of ESO474-G26, the central spheroidal component
dominates the light in the NIR bands, particularly in the Ks band,
while the rings emission becomes weaker from J to Ks bands: this
strongly suggests that most of the light relative to an old and
evolved stellar population comes from the central HG and in the rings
a stellar population as old as that in the HG is absent.

The integrated colors derived for ESO474-G26 are compared with those
of PRGs in the sample of \citet{Iod02a,Iod02b}, in order to check
whether there are differences in colors and average stellar
population age estimates between the main components of this galaxy
and other polar rings/disks, which could give some hints on the
formation mechanism. As explained in the previous studies on PRGs
(\citealt{Iod02a,Iod02b}), the B-K versus J-K diagram is used to
break the age-metallicity degeneracy; the J-K color is a good estimate
of the metallicity and it is quite insensitive to the presence of a
young stellar population.

The stellar population synthesis model by \citet{Bru03} were used to
reproduce the integrated colors in the selected regions, in order to
derive an estimate of the average (i.e. old plus the new bursts)
stellar population ages in the central component and in the ring-like
structures. We selected a set of models that were able to reproduce
the average integrated colors observed for the main components of
ESO474-G26.

The key input parameters for GISSEL ({\it Galaxies Isochrone Synthesis
  Spectral Evolution Library}, \citealt{Bru03}) are the Initial Mass
Function (IMF), the Star Formation Rate (SFR) and the metallicity.
For the central galaxy we adopted a star formation history with an
exponentially decreasing rate, that produces a reasonable fit of the
photometric properties of early-type galaxies in the local
Universe. It has the following analytical expression: $SFR(t) =
1/\tau\ exp (-t/\tau)$, where the $\tau$ parameter quantifies the
``time scale'' when the star formation was most efficient. Adopting
$\tau$ = 1 Gyr and $\tau$ = 7 Gyr, the correspondent evolutionary
tracks were derived for different metallicities (Z=0.1, Z=0.02,
Z=0.05, Z=0.008 and Z=0.0004), which were assumed constant with
age. For the ring-like structures of ESO474-G26 instead, since
  they have bluer colors than the host galaxy, which suggests even a
  younger age for this component, we used models with constant SFR
  computed for the same metallicities as above, because these models
  reproduce the integrated colors of local spiral galaxies, in which
  star formation is still active. In every model it has been assumed
that stars form according to the \citet{Sal55} IMF, in the range from
0.1 to 125 $M_{\odot}$.

\begin{figure*}
\centering
\includegraphics[width=15cm]{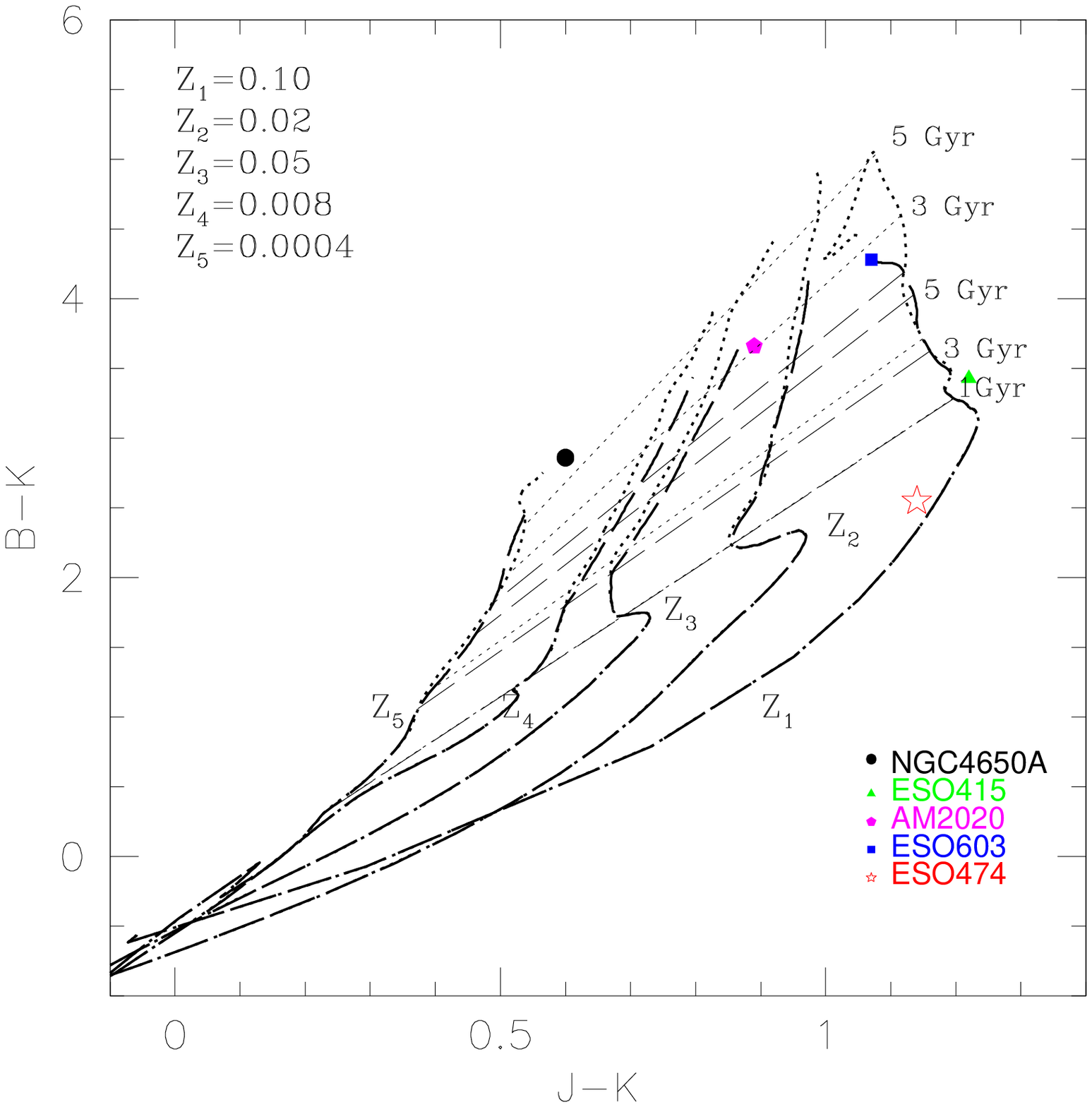}
\caption{B-K vs J-K diagram of the evolutionary tracks for the stellar
  synthesis models optimized for the central component of
  ESO474-G26. We used models with a characteristic time scale $\tau$ =
  1 Gyr (heavy dotted lines) and models with $\tau$ = 7 Gyrs (heavy
  dashed lines), computed for different metallicities as shown on the
  figure. Light dotted and and light dashed lines indicates loci of
  constant age for the different models; different ages are reported
  on the plot. The red star corresponds to the central object of
  ESO474-G26, while the other points correspond to the sample of PRGs
  in \citet{Iod02a, Iod02b}.} \label{ageHG}
\end{figure*}

\begin{figure*}
\centering
\includegraphics[width=15cm]{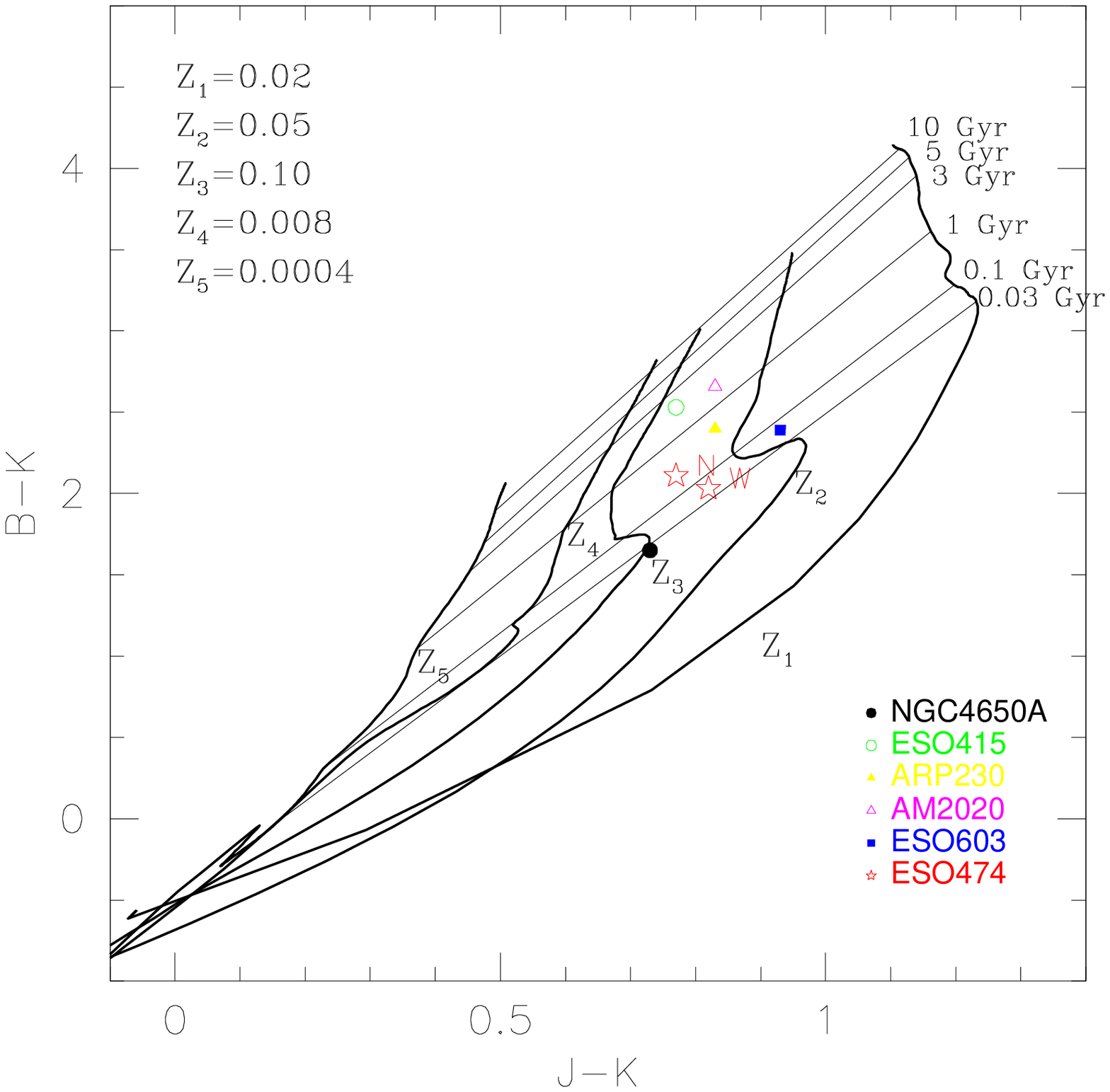}
\caption{B-K vs J-K diagram of the evolutionary tracks for the stellar
  synthesis models optimized for the central component of
  ESO474-G26. For the polar structures of ESO474-G26 we used models
  with constant SFR computed for different metallicities (heavy
  lines). Light lines are loci of constant age; different ages are
  quoted on the plot. Red stars correspond to different regions of
  the polar rings in ESO474-G26, while the other points correspond to
  the sample of PRGs in \citet{Iod02a, Iod02b}.} \label{agePR}
\end{figure*}

Figure \ref{ageHG} shows that the central galaxy is bluer in B-K color
than the average values for PRGs in the sample of \citet{Iod02a,
  Iod02b}, so a younger average age is to be expected, while figure
\ref{agePR} shows that the rings are bluer than the central
galaxy. The colors of the NW component are on comparable with
other narrow PRGs and not with NGC4650A, which is a wide polar disk,
while the SE side appear redder in J-K color (see Table
\ref{magnitudes}), but this could be the effect of a contamination due
to the presence of a very bright star in this direction; for this
reason the colors corresponding to the South and East side of the
rings are not reliable, and so they are not reported in Figure
\ref{agePR}. To account for the B-K and J-K colors the best model is
that obtained for Z = 0.1 for the central galaxy and Z = 0.05 for the
North and West side of rings, from which we derived an average age,
 of less than 1 Gyr for the
inner region and of less than 0.03 Gyrs for the outer ones. Such
values for ages and metallicity turn to be comparable with those
derived for the narrow PRGs in the sample of \citet{Iod02a, Iod02b}. This
metallicity values are inside the range observed for early type
galaxies ($Z_{4}\ \leq\ Z \leq\ Z_{1}$, \citealt{Both90}), but they
are higher with respect to the average value, which is around
$Z_{2}$. Furthermore, the J-K color for ESO474-G26 (and also for other
PRGs) are also consistent with the J-K colors derived by
\citet{Rossa07} for the interacting galaxies at an intermediate merger
stage ($0.5 \leq\ J-K \leq\ 1.5$): such range in Figures \ref{ageHG}
and \ref{agePR} correspond to metallicities between $Z_{3}$ and
$Z_{2}$.

We will show in section \ref{model} that, for both HG and rings,
the average ages derived by the integrated colors are comparable with
the epoch of formation for these galaxy components estimated by
simulations.

\subsection{2-Dimensional model of the host galaxy light distribution}\label{2D}

\begin{figure*}
\centering
\includegraphics[width=17cm]{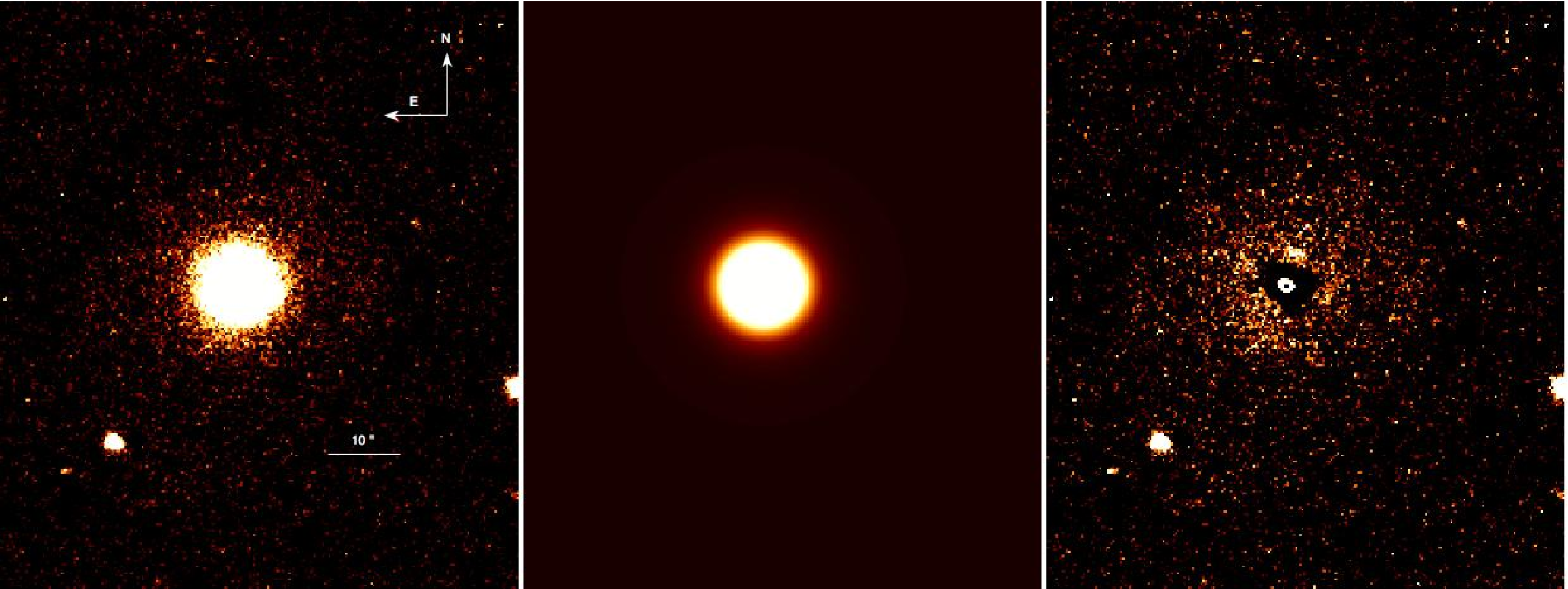}
\caption{2D fit of ESO474-G26. Left panel - K band image of
  ESO474-G26. Middle panel - K band model of the galaxy. Right panel -
  Residual of the subtraction of the model to the K band image.} \label{modK}
\end{figure*}

{\it Two-dimensional model}. We performed a 2-dimensional model of the
light distribution of the host galaxy in the Ks and in the B bands. To
this aim, we used the GALFIT task \citet{Peng02} and the resulting
structural parameters are listed in Table\ref{galfit}. The Ks image is
used to better constrain the structure of the central spheroidal
component, since it dominates the light in this band and the emission
from the rings and the dust absorption are very weak (see Table
\ref{galfit}). The best model for this component is obtained by
fitting the galaxy light through a single Sersic law
\citep{Ser68}. The result is shown in Figure \ref{modK}: except for
the center, where the residuals show the effect of the seeing, there
are no evident features, only a diffuse low-luminosity emission is
detectable in the North and SE directions, which is the weak residual
light coming from the rings.

\begin{figure*}
\centering
\includegraphics[width=17cm]{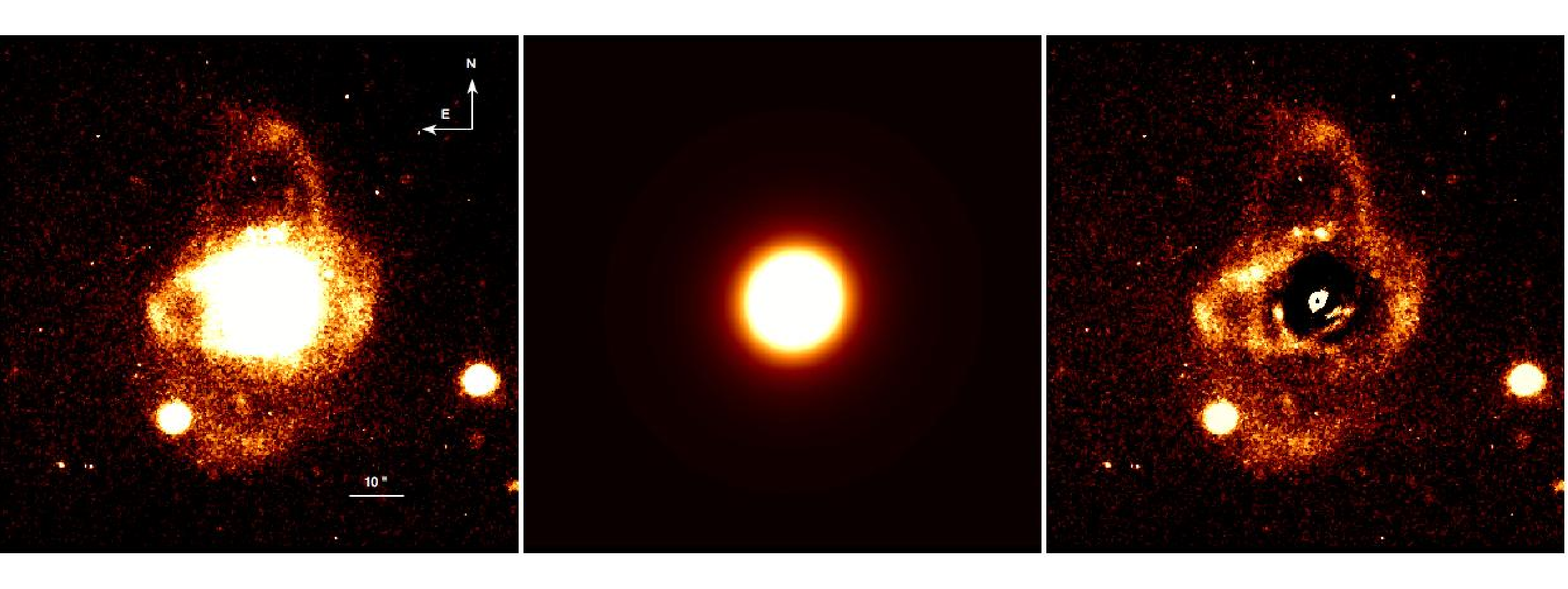}
\caption{2D fit of ESO474-G26. Left panel - B band image of
  ESO474-G26. Middle panel - B band model of the galaxy. Right panel -
  Residual of the subtraction of the model to the B band
  image.} \label{modB}
\end{figure*}

In order to analyze the complex ring-like structure in ESO474-G26,
which is very luminous in the B band, we have also derived the 2D
model in this band. By taking into account the constraints for the
central galaxy obtained by the 2D model in the Ks band and by
accurately mask all the ring structures (which need to be excluded by
the fit), the best 2D model in the B band is shown in Figure
\ref{modB}. The residuals appear very different from those in the Ks
band: the whole structures of both polar and equatorial rings stands
out very clear. Furthermore, it is evident another ring-like structure
in the very central regions: it approaches to the galaxy center in the
SW quadrant, it extends in the SE tracing a parabolic shape which
seems to be connected with the Northern arm of the polar ring.

\begin{figure*}
%\centering
\includegraphics[width=8cm]{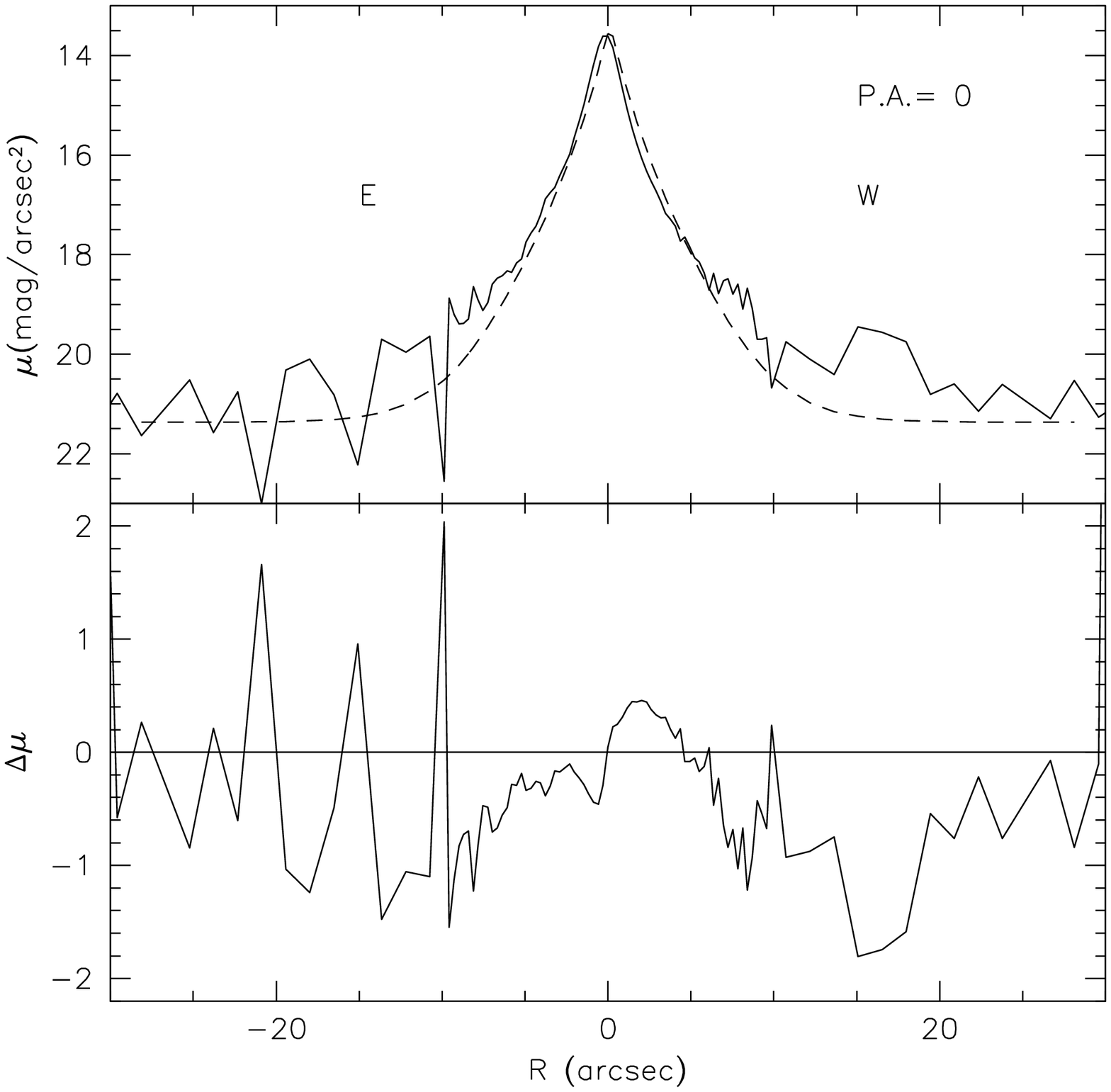}
\includegraphics[width=8cm]{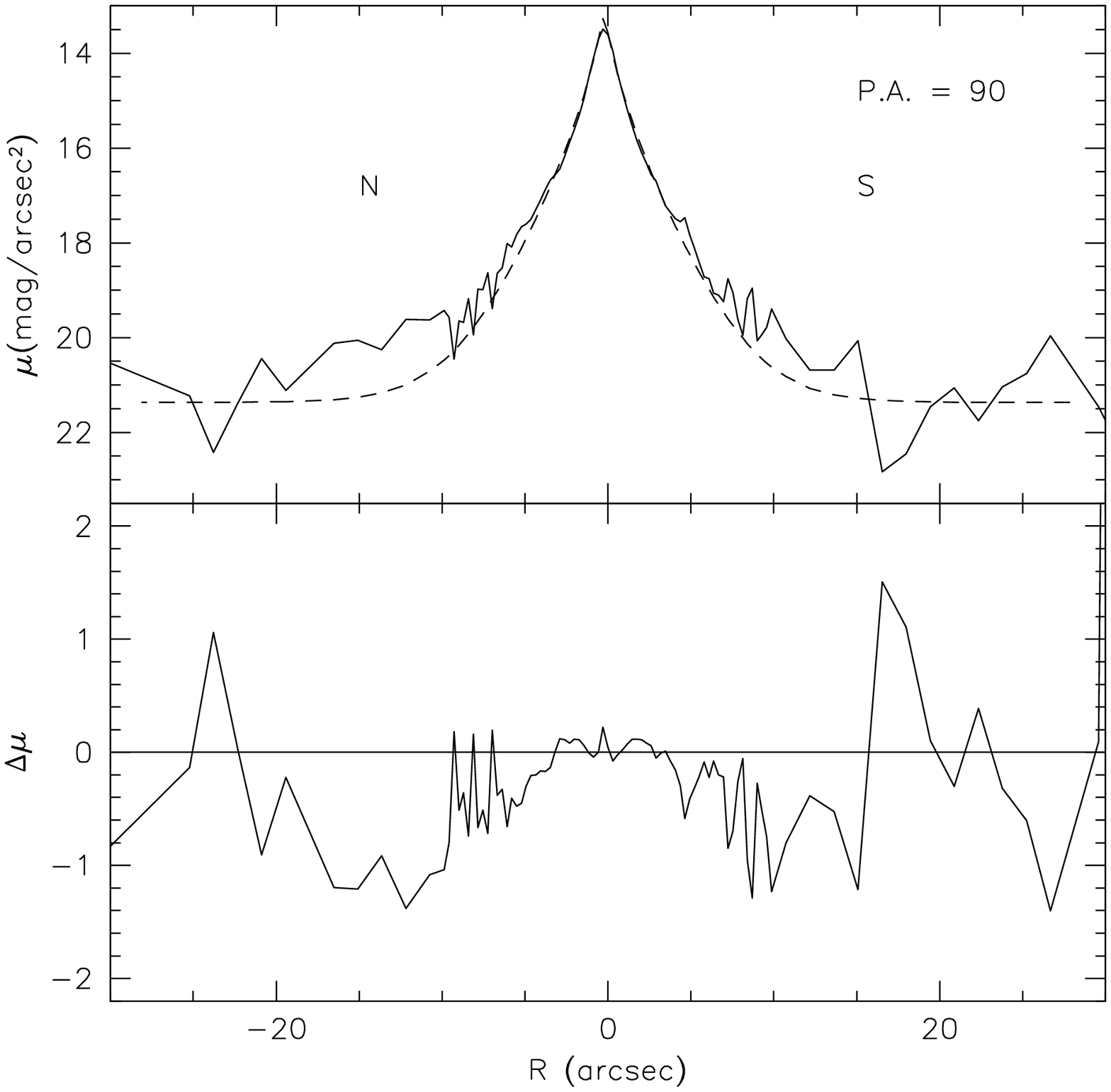}
\caption{Top left panel-2-D fit of ESO474-G26 light distribution in
  the K band. The observed light profile along the minor axis (EW), is
  compared with those derived by the fit (dashed line). Bottom left
  panel - Residuals between the observed and the fitted light
  profiles. Top right panel-The same as above but for major axis
  (NS). Bottom right panel - Residuals between the observed and the
  fitted light profiles.} \label{profmodK}
\end{figure*}

\begin{figure*}
%\centering
\includegraphics[width=8cm]{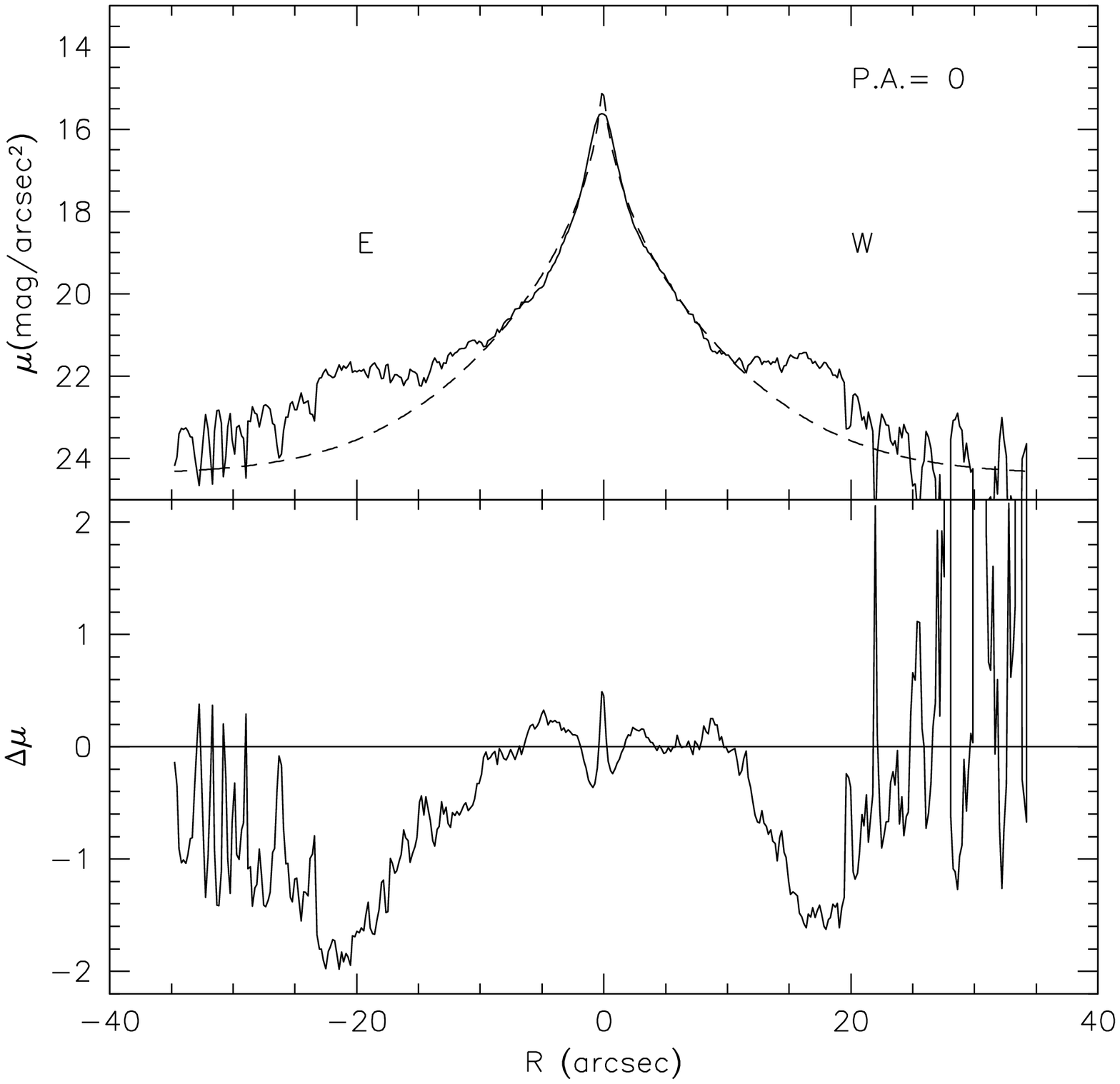}
\includegraphics[width=8cm]{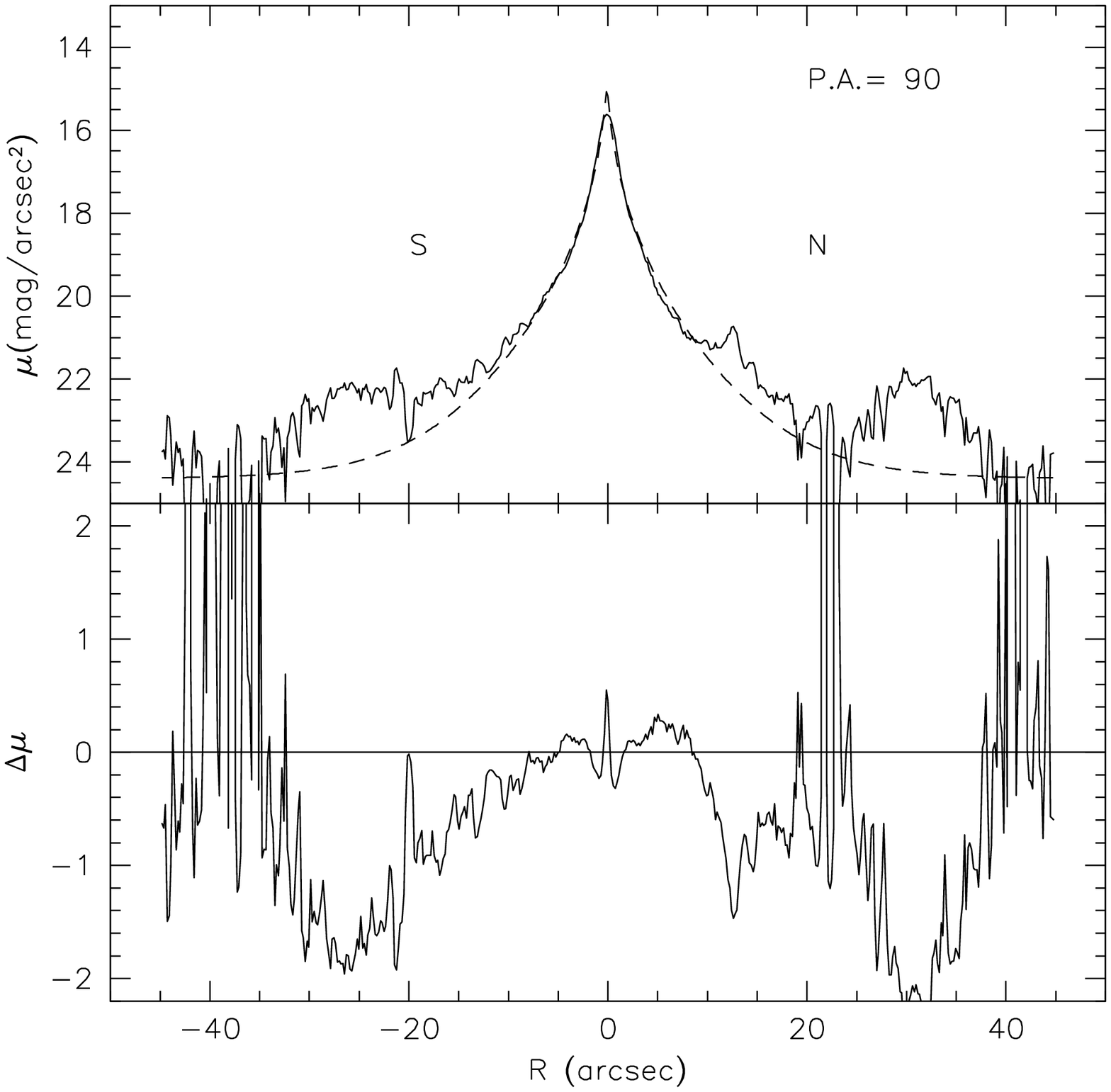}
\caption{Top left panel-2-D fit of ESO474-G26 light distribution in
  the B band. The observed light profile along the minor axis (EW), is
  compared with those derived by the fit (dashed line). Bottom left
  panel - Residuals between the observed and the fitted light
  profiles. Top right panel-The same as above but for major axis
  (NS). Bottom right panel - Residuals between the observed and the
  fitted light profiles.} \label{profmodB}
\end{figure*}

The comparison between the observed and fitted light profiles along
the galaxy major and minor axis (P.A.=0 and P.A.=90 degrees,
respectively) is shown in Figure \ref{profmodK} and in
Figure \ref{profmodB}, for Ks and B band respectively. In
all the profiles and at both position angles, the ``additional'' light
coming from the rings is evident at about 10 arcsec from the galaxy
center in the Ks band, and it is much more luminous and extended in
the B band, from 10 to 20 arcsec. Except for these regions, inside 10
arcsec, residuals are better than 0.2 mag. In the outer regions ($r
\ge 15$ arcsec in the Ks band and $r\ge 35$ arcsec in the B band)
light is dominated by background fluctuations.

\begin{table*}
\caption{\label{galfit}Galfit parameters.} \centering
\begin{tabular}{lcccc}
\hline\hline
Parameter&Value&Parameter&Value\\
\hline
{\it K band}& &{\it B band}&\\
\hline
Component & Sersic&Component & Sersic\\
Integrated magnitude& 11.30&Integrated magnitude &17.78\\
Effective radius& $7 \pm 0.01$ arcsec&Effective radius & $15.52 \pm 0.05$ arcsec\\
Sersic index & $1.52 \pm 0.01$&Sersic index & $3.13 \pm 0.01$\\
Axis ratio (b/a)& $0.97 \pm 0.01$&Axis ratio (b/a)& $0.99 \pm 0.01$\\
Position Angle & $81.3 \pm 1.7$&Position Angle & $101.0 \pm 5.6$\\
\hline
\end{tabular}
%\tablefoot{ \tablefoottext{a}{NASA/IPAC Extragalactic Database}}
\end{table*}

\section{Comparing the data with models}\label{model}
To address the question on the formation scenario for ESO~474-G26 we
attempt to best fit its overall SEDs and global properties, analyzing
a large set of SPH simulations.

Our SPH simulations of galaxy formation and evolution start from the
same initial conditions described in \cite{MC03} (MC03 hereafter)
  and \cite{Maz03} (and references therein) i.e., collapsing triaxial
  systems composed of dark matter (DM) and gas with density
  distribution $\rho \propto$r$^{-1}$, in different proportions and
different total masses. We then allowed a large set of galaxy
encounters involving systems with a range of mass ratios from 1:1 to
1:10.
%In the following minor mergers are excluded since they do not
%represent the observed global properties.\\
In order to exploit a vast range of orbital parameters, we carried out
different simulations for each pair of interacting systems, varying
the orbital initial conditions in order to have, for the ideal
Keplerian orbit of two mass points, the first peri-center separation,
p, ranging from the initial length of the major axis of the dark
matter triaxial halo of the primary system to 1/10 of the same (major)
axis.\\
For each of these separations, we changed the eccentricity in order to
have hyperbolic orbits of different energies. For the most part we
studied direct encounters, where the spins of the systems are equal
(MC03), generally parallel to each other, and perpendicular to the
orbital plane.  However, we also analyzed some cases with misaligned
spins in order to enhance the effects of the system initial rotation
on the results. Moreover, for a given set of encounters with the same
orbital parameters we also examined the role of increasing initial gas
fractions.\\
%These simulations will be fully discussed elsewhere (Mazzei, in prep).
All simulations include self--gravity of gas, stars and DM, radiative
cooling based on standard cooling function including metals,
hydrodynamical pressure, shock heating, artificial viscosity, star
formation (SF) and feedback from evolved stars and type II SNe, and
chemical enrichment. The minimum temperature reached is
  $10^{3}$\,K to save time computing. Here, as in the following, cold
  gas is gas with temperature lower than $10^{4}$\,K, given that its
  cooling timescale is shorter than the snapshot time-range, as we
  will point out later on. The gravitational softening is taken to be
  0.5 and 1\,Kpc respectively for the gas and DM particles. The
Initial Mass Function (IMF) is of Salpeter's type with upper mass
limit of 100$\,M_\odot$ and lower mass limit of 0.01$\,M_\odot$
(\citealt{Sal55}; see MC03 and references therein for a discussion).
All our simulations provide the synthetic SED at each evolutionary
step. The SED accounts for chemical evolution, stellar emission,
internal extinction and re-emission by dust in a self-consistent way,
as described in \citet{Spav09} and references therein, and extends
over almost four orders of magnitude in wavelength, i.e., from 0.05 to
1000 $\mu$m. So, each simulation self-consistently accounts for
morphological, dynamical and chemo-photometric evolution.

The whole SED of ESO~474-G26 and its global properties are well
matched by a major merger, i.e. with a 1:1 mass ratio. The total
initial mass of the systems is 4$\times$10$^{12}$ M$_\odot$ with gas
fraction 0.10, so that the total initial mass of the gas is
4$\times$10$^{11}$ M$_\odot$. The mass particle resolution is
1.33$\times$10$^{7}$ M$_\odot$ for gas and 1.2$\times$10$^{8}$
M$_\odot$ for DM particles. This requires 60000 initial
particles.\\ The first pericenter separation, 101.4~Kpc, corresponds
to 1/10 of the major axis of their halo; the orbit eccentricity is 1.3
and the anomaly corresponds to 200 degrees. The initial haloes have
perpendicular spins ($\lambda$ = 0.058) and triaxiality ratio 0.84
\citep{MC03}. Stars are born in the inner regions of their halos after
about 2.5 Gyr from the beginning. Galaxies grow changing their shapes
step by step as their trajectories are approaching and their halos
mixing.

\subsection{Evolution of ESO~474-G26}

\begin{figure}
%  \centering
\includegraphics[width=9cm]{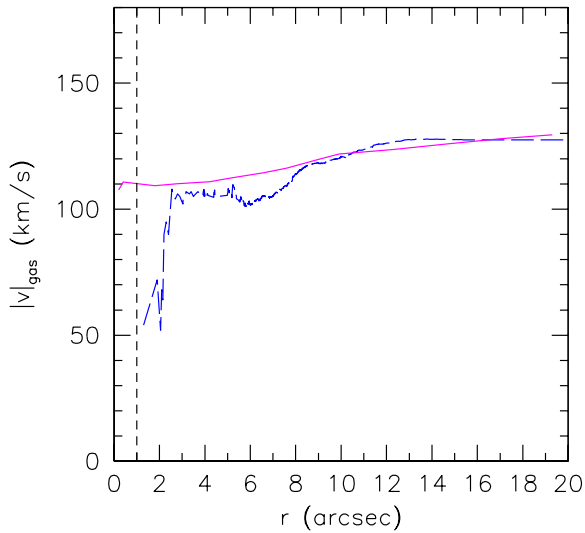}
\caption{Blu dashed line shows the mean velocity distribution predicted for
  the cold and warm gas in our model (see text) along the galaxy major
  axis. Magenta solid line corresponds to the averaged velocity distribution
  as derived from H$\alpha$ measurements along the galaxy major axis,
  corrected for the inclination. Vertical dashed line corresponds to
  the inner region (R$\le$1\,Kpc) where model sampling could affect
  the results.}
       \label{Fig_starvelESO474_G26}
   \end{figure}

\begin{figure}
%  \centering
\includegraphics[width=9cm]{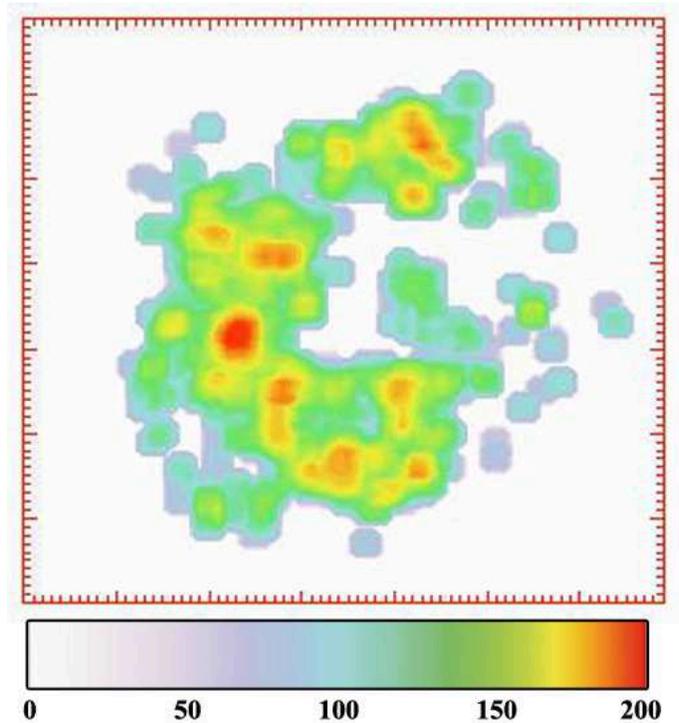}
\caption{YZ map, 30\,Kpc $\times$ 30\,Kpc, of cold and warm gas (see text)
  at the snapshot corresponding to the final merging stage ($\sim$ 9
  Gyr). The mass of cold and warm gas is normalized to its total mass
    within the map; its density contrast is 200 with 60 equispaced
    levels and spatial resolution 0.4\,Kpc.}
       \label{gas}
   \end{figure}

We match the global properties of ESO~474-G26 during a very active
phase, due to the strong interaction between the systems, i.e. 11.2
Gyrs from the beginning of the simulation.\\ The older stellar
populations are 8.5 Gyr old, however the large majority of star
clusters are younger than 6.5 Gyr. By averaging population ages
inside $\sim$r$_{eff}$(B) and $\sim$4r$_{eff}$(B), our fit corresponds
respectively to a galaxy age of 2~Gyr and 4~Gyr. The B band absolute
magnitude of the model, -22.31 mag, agrees well with the
cosmologically corrected value from HyperLeda and the value in Table
\ref{global}, as so as the K band absolute of the model, -25.32 (see
Table \ref{global}). The mean velocity distribution along the galaxy
major axis of cold and warm gas, i.e. gas with temperature lower than
  $10^{4}$\,K whose cooling timescale is shorter than the snapshot
  time-range (38Myr), compares with the same distribution derived
from H$\alpha$ measurements at the major axis position angles (P.A) of
\citet{Whi90}, as shown in Figure \ref{Fig_starvelESO474_G26};
measurements in \citet{Whi90} are corrected to account for a line of
sight inclination angle of 52.2 degree (HyperLeda). Vertical dashed
line corresponds to the inner region (R$\le$1\ Kpc) where model
sampling could affect the results. We point out that the slightly
difference, of almost 10 km/s, can be accounted for the error in the
recession velocity of the galaxy (12 Km/s, by HyperLeda).
%and the uncertainty in the line of sight inclination angle.
At the selected snapshot the cold and warm gas, which will form
new stars, is distributed in a ring (see Figure \ref{gas}) surrounding
the central spheroid.

\begin{figure}
 % \centering
\includegraphics[width=9cm]{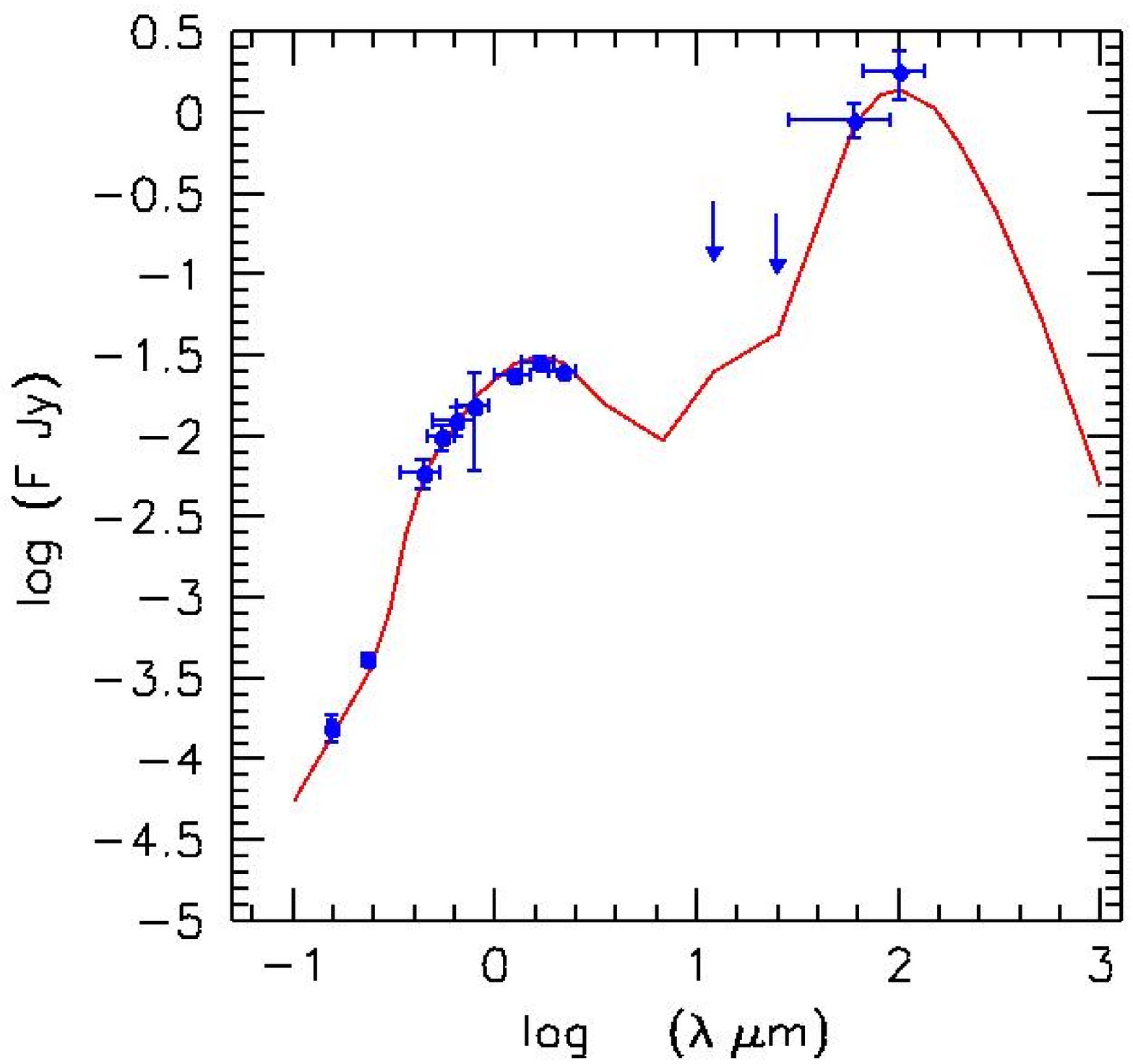}
\caption{Continuous line (red) show the prediction of our model (see
  text). (Blue) filled circles correspond to data from NED and from
  Table \ref{global}. Arrows show upper limits; error bars account for band width
  and 3$\sigma$ uncertainties. Dust components (warm and cold with
  PAH as discussed in \citet{Maz92} with the same average
  properties as derived by \citet{MaeDe94} for a complete
  sample of nearby early-type galaxies are included
  (i.e. I$_0$=46I$_{local}$, Iw=110I$_{local}$, R$_{wc}$=0.27 and
  r$_d$=30r$_c$, as in \citealt{Maetal94})}.
       \label{Fig_sedESO474_G26}
   \end{figure}

Figure \ref {Fig_sedESO474_G26} shows the best fit of the predicted SED
by the simulation for ESO474-G26 to the available data.  Dust
components, warm and cold with PAH as discussed in \citet{Maz92} with
the same average properties as derived by \citet{Maetal94} and
\citet{MaeDe94} for a complete sample of nearby early-type galaxies
are included in the far-IR SED. The SFR inside 6 r$_{eff}$(B) ($\sim$
60 Kpc) is 41 $M_{\odot}/yr$, in agreement with radio estimates (see
Sec. \ref{radio}). The total mass inside r$_{eff}$(B) is
1.30$\times$10$^{11}\,M_\odot$ with 14.4\% of dark matter, inside
2r$_{eff}$(B) raises to 3.50$\times$10$^{11}\,M_\odot$ with 25\% of DM
and to 4.650$\times$10$^{11}\,M_\odot$ with 44\% of DM inside
4$_{eff}$(B). The predicted M$_{tot}$/L$_{B}$ ratio goes from
3\,M$_\odot$/L$_\odot$ at r$_{eff}$(B) to 6.5\,M$_\odot$/L$_\odot$ at
4r$_{eff}$(B).

\begin{figure}
%\centering
\includegraphics[width=9cm]{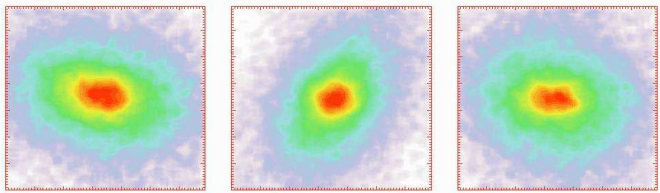}
\caption{B band intrinsic flux map, 60\,Kpc $\times$ 60\,Kpc,
  of XY, YZ, and XZ projections at the snapshot corresponding to 0.5
  Gyr after the merging. The remnant becomes smooth and the morphology
  resembles that of spheroidal early-type system. Each panel is
    normalized to its total flux; the density contrast is 200 with 60
    equispaced levels, the spatial resolution
    1.2\,Kpc. The color scale is the same as in Fig. \ref{gas}.}\label{0.5Gyr}
\end{figure}

By inspecting the simulation at the snapshot corresponding to 0.5 Gyr
after the merging (Figure \ref{0.5Gyr}), the remnant becomes
smooth and the morphology resembles that of spheroidal early-type
system.

\section{Discussion and conclusions}\label{conc}
We have presented a detailed photometric study of the double ringed
galaxy ESO474-G26, based on new NIR observations. The main results of
the present work are: {\it i)} the central spheroidal component
dominates the light in the NIR bands, particularly in the Ks band,
while the rings emission becomes weaker from J to Ks bands; {\it ii)}
by using the stellar population synthesis model (see sec. \ref{age}), the last
burst of star formation is dated to be less than 1 Gyr for the central
galaxy, and between 0.1 and 0.03 Gyr for the ring-like structures,
comparable with those of other PRGs with narrow rings (see Figure \ref{ageHG} and
\ref{agePR}); {\it iii)} the 2D model of the light distribution suggests that
one of the two rings, most probably the polar one, extends till the
galaxy center (see Figure \ref{modB}); {\it iv)} the SED turns to be well
matched by a major merger (see Figure \ref {Fig_sedESO474_G26}).

The main goal of the analysis presented in this work is to address the
most reliable formation scenario for ESO474-G26, by reconciling the
observed properties for this peculiar object with those predicted by
different formation mechanisms for such kind of systems. As widely
discussed into sec. \ref{intro}, the up-to-date scenarios proposed for
PRGs formation are (1) the disruption of a dwarf companion galaxy
orbiting around an early-type system, or the tidal accretion of gas
stripping from a disk galaxy outskirts, (2) a dissipative merging of
two disk galaxies, (3) accretion of cold gas from cosmic web
filaments. In the case of ESO474-G26, we expect that the most likely
formation scenario has to account for {\it i)} the morphology,
i.e. the presence of two orthogonal ring-like structures around an
almost spherical early-type object, {\it ii)} the observed colors and
ages for both components, i.e. a good fit for the observed SED; {\it
  iii)} the observed kinematics; {\it iv)} the gas content and
distribution. To this aim, we used the infrared photometry which has
the main advantages to be less affected by dust absorption so it well
traces the older stellar population.

In order to reproduce the structure of ESO474-G26, \citet{Res05} have
performed N-body simulations of a low-velocity head-on collision
between two galaxies with orthogonal spiral disks. In their
simulations, the main galaxy was a giant spiral galaxy with a low gas
fraction, while the second galaxy was less massive but with a higher
gas content. They found that, 350 Myr after the first crossing, the
merger remnant shows an equatorial ring made up of stars coming from
the intruder galaxy, and an expanding collisional polar ring. They
found that the polar ring is a transient feature and that 1 Gyr after
the full merging, the remnant becomes an elliptical-like object.

On the basis of these previous results and by taking into account the
new NIR photometry, we have developed new SPH simulations of galaxy
formation and evolution (Sect.\ref{model}), in order to test the major
merger scenario for ESO474-G26. Differently from \citet{Res05},
  in our simulations no galaxies exist at the beginning, but only
  collapsing triaxial halos composed of dark matter and gas, in equal
  proportions in both the systems, with equal total masses (1:1) and
  perpendicular spins (Sect.\ref{model} for details). In the paper of
  \citet{Res05} galaxies with mass ratio 2.5:1 are separated by
  70\,Kpc and their relative velocity was 70 km/s whereas, in our
  case, the initial relative velocity of their centre of mass is
  $\simeq$104\,Km/s and its position corresponds to 233\,Kpc in the
  XY, orbital plane. The total inital mass in \citet{Res05} is
  8.8$\times$10$^{11}$\,$M_{\odot}$, while in our simulation it is 4.5
  times more.  We confirm that a major merger of two haloes with a
1:1 mass ratio is able to well reproduce the observed structure and
the properties of ESO474-G26: in particular, {\it i)} an excellent fit
of the SED (see Figure \ref{Fig_sedESO474_G26}), which is
well-constrained by the NIR fluxes; by inspecting the simulation at
the snapshot corresponding to 0.5 Gyr after the merging, {\it ii)} the
remnant becomes smooth, the ring-like structure vanishes and the
morphology resembles that of spheroidal early-type system (see Figure
\ref{0.5Gyr}); {\it iii)} the predicted gas kinematics is consistent
with the observed one along the galaxy major axis (see Figure
\ref{Fig_starvelESO474_G26}); {\it iii)} the cold gas, which will form
new stars, is distributed in a ring (see Figure \ref{gas}) surrounding
the central spheroid.  As already pointed out by \citet{Res05}, also
this new simulation suggests that the structure of ESO474-G26 could be
a transient phase and not a stable dynamical configuration.

In order to complete our analysis on the evolution history of
ESO474-G26, we now examine if the other possible formation scenarios,
could also equally well match its observed properties.\\ The tidal
accretion scenario, in which gas is stripped from a gas-rich donor in
a particular orbital configuration \citep{Bou03}, is able
to produce wide rings and/or disks both around a disk or an elliptical
galaxy: in this framework, in the field around the new forming ring
galaxy the gas-rich donor galaxy is still present. In the case of
ESO474-G26, inside a radius of about five times its diameter, as
suggested by \citet{Broc97}, there are no close companions as
possible donor galaxy candidates. Moreover, in the tidal accretion
scenario, the total amount of accreted gas by the early-type object is
about 10\% of the gas in the disk donor galaxy, i.e. up to $10^{9} M_{\odot}$:
the high baryonic mass (star plus gas) in the rings of ESO474-G26,
which is about $10^{10} M_{\odot}$, turns to be not consistent with this limit
and further confirms that the tidal accretion need to be ruled out.
The gradual disruption of a dwarf satellite galaxy cannot reconcile
with a multiple ring structure.

Finally, the cold accretion scenario predict the formation of wide
disk-like structures around an host galaxy. In this scenario, a
long-lived polar structure may form through cold gas accretion along a
filament extended for $\sim 1$ Mpc into the virialized dark matter
halo \citep{Mac06}. In this formation scenario, there are no limits to
the mass of the accreted material, thus, a very massive polar disk may
develop either around a stellar disk or a spheroid. \citet{Brook08},
by using high-resolution cosmological simulations of galaxy formation,
have confirmed and strengthened the formation scenario proposed by
\citet{Mac06}. However, \citet{Brook08} in their study have referred
to objects characterized by polar structures with disk galaxy
characteristics. In other galaxies, such as ESO474-G26, the polar
structure is better described as a ring, with gas and stars in a
narrow annulus. For this reason, the study of \citet{Brook08} is
not conclusive on this issue, and the classic merging and accretion
models remain viable explanations for the formation of narrow polar
ring structures. \citet{Spav11}, by studying the formation
mechanism of the PRG UGC7576, were able to test and confirm the cold
accretion for this object, whose polar structure is a ring rather than
a disk. However, differently from ESO474-G26, UGC7576 have a wide
ring-like structure. Moreover, the narrow rings observed in ESO474-G26
are very faint structures, and they are not able to survive to the
repeated accretion and mergers predicted by \citet{Brook08}.

We can thus conclude that the major merger is the most plausible
scenario for the formation of the complex double ringed structure of
ESO474-G26.

\section*{Acknowledgements}
The authors thank the referee, Frederic Bournaud, for the detailed and
constructive report, which allowed them to improve the paper.
E.I. wish to thank E. Pompei for the support given during the data
acquisition. This work is based on observations made with ESO
Telescopes at the Paranal Observatories under programme ID $<$ 70.B -
0253(A) $>$ and $<$ 74.B - 0626(A) $>$. We acknowledge financial
contribution from the agreement ASI-INAF I/009/10/0 and
M.S. acknowledges financial support from the ``Fondi di Ateneo 2011''
(ex 60 \%) of Padua University. V.R. acknowledges partial financial
support from the RFBR grant $11-02-00471-$a.

\bibliography{bibliografia}

\bsp

\label{lastpage}

\end{document}